\documentclass[a4paper,10pt]{article}
\usepackage[utf8]{inputenc}
\usepackage{amsmath}
\usepackage{amsfonts}
\usepackage{xcolor}
\usepackage{booktabs}
\usepackage{array}
\usepackage{graphicx}
\usepackage[colorlinks=True,allcolors=blue]{hyperref}
\usepackage{authblk}
\usepackage[left=0.7in,right=0.7in,top=1in,bottom=0.7in]{geometry} 
\usepackage{caption}  
\title{Multipolar social systems:\\Measuring polarization beyond dichotomous contexts}

\author[1,2]{Samuel Martin-Gutierrez}
\author[1]{Juan C. Losada}
\author[1,*]{Rosa M. Benito}
\affil[1]{Grupo de Sistemas Complejos,
Escuela T\'ecnica Superior de Ingenier\'ia Agron\'omica, Alimentaria y de Biosistemas, 
Universidad Polit\'ecnica de Madrid, 
Av. Puerta de Hierro 2-4, 28040 Madrid, Spain}
\affil[2]{Networked Inequality Group, Complexity Science Hub Vienna, Josefstaedter Str. 39, 1170 Wien, Austria}
\affil[*]{rosamaria.benito@upm.es}

\date{}

\begin{document}



\maketitle

\begin{abstract}
Social polarization is a growing concern worldwide, as it strains social relations, erodes trust in institutions, and thus threatens democratic societies. Academic efforts to understand this phenomenon have traditionally approached it from a one-dimensional perspective, focusing on bipolar or dichotomous systems.  However, political conflicts often involve not only two, but multiple potentially dissenting factions. The most representative examples are multi-party democracies, where the multilateral tensions among different parties often lead to gridlock and uncertainty. Despite the prevalence of these multipolar systems, there is still a lack of suitable analytical tools to study their intricate polarization patterns. In this work, we develop an analytical framework consisting of an inherently multipolar model for unbiased ideological spaces, a method to infer multidimensional opinions from interaction networks, and novel multidimensional polarization metrics that quantify several aspects of ideological polarization and bring new insights into the analysis of high-dimensional opinion distributions. Crucially, our multidimensional framework does not assume the underlying ideological structure, such as conservative vs progressive, liberal vs authoritarian, etc. Instead, it reveals the natural space that best describes the social landscape, which does not necessarily correspond to traditional categories. We illustrate the application of this framework in quadripolar and pentapolar real-world democratic processes, finding non-trivial ideological structures with clear connections to the underlying social context. Our methodology offers a comprehensive perspective of multilateral social tensions, as it incorporates complementary aspects of polarization: network segregation, opinion extremeness, and issue alignment.
\end{abstract}


\section*{Introduction}

Social polarization is a pervasive phenomenon \cite{doi:10.1073/pnas.2116950118} that has been observed in a wide variety of contexts like elections \cite{groverPolarizationAcculturationUS2019,olivaresOpinionPolarizationDichotomous2019}, referenda \cite{hoboltBrexitVoteDivided2016} and around controversial issues \cite{gollwitzerPartisanDifferencesPhysical2020,atienza-barthelemyRelationshipIdeologyLanguage2019,antonioUnbearableLightnessPolitics2011,PhysRevLett.110.088701}. It has been argued that polarization exerts a negative influence over democracy \cite{doi:10.1073/pnas.2116863118}, undermining social and economic relationships \cite{iyengarOriginsConsequencesAffective2019,doi:10.1073/pnas.2102140118}, causing legislative gridlocks \cite{finkelPoliticalSectarianismAmerica2020,bondPresidentialCongressionalRelationsEra2015}, and even posing health risks \cite{doi:10.1073/pnas.2102144118,gollwitzerPartisanDifferencesPhysical2020,bavelUsingSocialBehavioural2020}. However, it has also been found that some level of polarization can have healthy effects, as it produces clearer party choices \cite{daltonQuantityQualityParty2008,levenduskyClearerCuesMore2010} and more constructive, focused and linguistically diverse debates \cite{shiWisdomPolarizedCrowds2019}. In any case, to fully understand social dynamics we need tools to detect and measure polarization \cite{wallerQuantifyingSocialOrganization2021}, as it is a core feature of many systems. In this work we develop a multidimensional opinion inference framework for networked systems and tools to measure and characterize the polarization patterns of the inferred opinion distributions.

Polarization is usually defined as the division of a population into two conflicting or contrasting groups \cite{schweighoferWeightedBalanceModel2020,guerraMeasurePolarizationSocial2013}. This intuitive notion coincides with one of the aspects of polarization: {\it opinion extremeness}, which is quantified by measuring how concentrated is the distribution of attitudes of the population towards some issue at two polar opposite positions (for example, immigration, abortion, or cannabis legalization). 
Although one-dimensional (or bipolar) aspects of polarization like opinion extremeness are by far the most studied \cite{bramsonUnderstandingPolarizationMeanings2016,dimaggioHaveAmericanSocial1996}, 
there are other fundamental facets that can only be understood from a multidimensional perspective \cite{schweighoferWeightedBalanceModel2020,baldassarriPartisansConstraintPolitical2008,converseNatureBeliefSystems1964}.
For instance, in a polarized context the policy preferences of the population on different issues are often correlated (a person that is in favor of strong immigration controls may also be against free abortion and cannabis legalization, and vice versa). 
This means that the individuals' opinions get aligned along multiple lines of potential disagreement, exacerbating tensions and leading to a 
new kind of polarization: \emph{issue constraint} \cite{doi:10.1073/pnas.2102148118,mccoyPolarizationGlobalCrisis2018}.
If a system is highly polarized in the sense of issue constraint, one can infer the ideological positions of an individual towards several issues by knowing her position on only one of them. A typical example is a highly partisan democracy \cite{baldassarriPartisansConstraintPolitical2008,fiorinaDisconnectBreakdownRepresentation2009}, where party identification determines many of the ideological positions of individuals. Societies with strong partisanship can also present \emph{affective polarization} \cite{iyengarOriginsConsequencesAffective2019}, meaning that supporters of the different parties dislike or even loathe each other.

Issue constraint is usually measured by analyzing the correlation between one-dimensional 
attitude distributions (one per issue or topic) built from opinion polls. These polls ask dichotomous questions about topics such as public health insurance, treatment of minorities or abortion \cite{baldassarriPartisansConstraintPolitical2008}. The answers are typically organized in a Likert-like scale, ranging from complete agreement with something to complete disagreement. While this approach offers a multidimensional perspective, it is still fundamentally bipolar, as it consists in a combination of dichotomous issues. However, there are social contexts whose ideological space can not be modeled as a combination of bipolar axes: scenarios where opinions about a given issue are divided into more than two qualitatively different positions \cite{reinaModelBestofNNestsite2017,siegenfeldNegativeRepresentationInstability2020,franciModelindependentTheoryConsensus2020,reverdyDynamicalValuebasedDecision2020,bizyaevaGeneralModelOpinion2020}. We call these contexts \emph{multipolar systems}. The most representative examples are multi-party democracies, in which more than two parties have a realistic chance of obtaining significant representation. We model these systems by considering each party as an opinion pole. The intrinsic complexity of reaching multilateral government agreements sometimes leads to political instability and gridlock, causing recurrent repetition of elections and long periods of interim executives. Recent examples are found in Belgium \cite{BelgiumAgreesGovernment2020}, Israel \cite{serhanWhatPointRepeat2019}, or Spain \cite{SpainUpperHouse2020}. Understanding \emph{multipolar systems} \cite{PhysRevE.104.034303} is critical to identify the main lines of disagreement 
and address these issues, but their multidimensional polarization patterns are still largely unexplored.
In this paper we model the ideological space of multipolar systems with $n$ opinion poles by placing each pole (for example, the parties in a multi-party democracy) at the vertex of a regular simplex of dimension $n-1$ (a multidimensional generalization of an equilateral triangle, which would correspond to a tripolar system). This way every pole is at the same distance of the others, avoiding the introduction of a priori biases. 


Polarization is also closely linked to the way individuals from different social groups interact, so to get a complete understanding of polarization, it is necessary to consider not only the ideological stances of the polarized set of individuals or parties, but also the relationships among them \cite{garciaIdeologicalTemporalComponents2015,baldassarriDynamicsPoliticalPolarization,Conover2011PoliticalPO}. After all, the formation of social ties between individuals is often related to their ideological affinity \cite{flacheSmallWorldsCultural2011}, potentially leading to {\it network polarization} \cite{doi:10.1073/pnas.2102147118,doi:10.1073/pnas.2102141118}; that is, the organization of a social network in highly connected subgroups (or clusters) with weak inter-group connectivity \cite{garimellaQuantifyingControversySocial2018,guerraMeasurePolarizationSocial2013,doi:10.1126/sciadv.aax7310}.
Taking this into account, we use an opinion inference framework based on networks of user interactions in online social media \cite{moralesMeasuringPoliticalPolarization2015} (although the framework can be applied to any networked system). This methodology allows us to estimate the opinion distribution of the users and circumvents the common drawbacks of attitude inference techniques based on the content of user messages, such as the dependence on context and language \cite{guerraMeasurePolarizationSocial2013}.

To characterize and measure the polarization of the inferred opinion distribution we propose different metrics based on the covariance matrix,  
which is the multidimensional generalization of the variance, a quantity often adopted as a one-dimensional measure of polarization \cite{bramsonUnderstandingPolarizationMeanings2016,dimaggioHaveAmericanSocial1996}. In particular, we use the trace of the covariance matrix (the total variation) as a global measure of {\it opinion extremeness}, and its eigendecomposition to quantify {\it pole constraint} (a multipolar analogue of issue constraint), obtaining the direction of maximum polarization in the ideological space \cite{barberaTweetingLeftRight2015a}.

Summarizing, our multipolar framework is composed of three main elements: a multipolar ideological space, an opinion inference technique based on user interactions that yields a multidimensional opinion distribution and a set of polarization measures that are applied to that distribution. This approach synthesizes the ideological and networked perspectives of polarization and provides means to study multipolar analogues of both opinion extremeness and issue constraint.  To illustrate our new framework in practice, we apply it to Twitter data of two multipolar electoral contexts, the Spanish general elections of the 20 of December 2015, with four poles, and the Spanish general elections of the 28 of April 2019, with five poles.

\section*{Results}

We will begin this section by laying out the foundations of the multidimensional opinion inference methodology, the ideological space built from pole vectors and the multi-polarization measures we will use on the opinion distributions. Then, we will apply this framework to two empirical systems, showing how our results can be easily interpreted in terms of the underlying social context and illustrating the usefulness of this methodology to characterize multipolar systems. To help build an intuition about the methodological framework and facilitate the analysis of the data, we have prepared interactive visualizations that can be found at \url{https://vis.csh.ac.at/multipolar-viz}.

\subsection*{Opinion inference and polarization measurement}

\subsubsection*{The multidimensional opinion model}
\label{sec:fj_model}

Our multidimensional opinion inference technique is a generalization of a bipolar (one-dimensional) methodology \cite{moralesMeasuringPoliticalPolarization2015} based on models of opinion dynamics. But instead of using a model to understand how opinions are disseminated and adopted \cite{sayamaEnhancedAbilityInformation2020,bottcherCompetingActivistsPolitical2020}, we use it as an inference tool. The process consists in building a network of social interactions from empirical data, identifying the opinion leaders and their respective ideological positions. Then, we use the model to propagate the leaders' opinions throughout the rest of the nodes. Finally, we take the model's outputs (the converged opinions) as the inferred opinions of the nodes. Therefore, the opinion inference technique requires:



\begin{itemize}
	\item \emph{A network of interactions}: we build it using data from social networking sites.
	\item \emph{The elite}: $n$ small disjoint subsets of nodes assigned to each of the $n$ opinion poles to be used as opinion seeds.  The nodes of each subset should have different and extreme opinions about the topic under study. We will choose them using contextual information.
	\item \emph{An opinion learning mechanism}: An iterative opinion updating procedure to infer the opinions of the rest of the nodes (\textit{the listeners}). 
\end{itemize}

For the learning mechanism, we have adopted a modified version of the DeGroot linear learning model \cite{degrootReachingConsensus1974}. In this model, each \emph{elite} node is assigned a fixed opinion depending on the option they support. Then, the opinion of the remaining nodes of the network (the \emph{listeners}) is computed by iteratively averaging the opinions of their neighbors. This modified version of the DeGroot model including nodes with fixed opinions
was originally developed by Friedkin and Johnsen \cite{Friedkin1999} and has been extensively applied in bipolar contexts \cite{moralesMeasuringPoliticalPolarization2015,atienza-barthelemyRelationshipIdeologyLanguage2019,olivaresOpinionPolarizationDichotomous2019}.
A multidimensional extension of the Friedkin-Johnsen (FJ) model has been proposed and thoroughly characterized by Parsegov et al. \cite{parsegovNovelMultidimensionalModels2017}, so we will draw on their results for our inference technique.


In the FJ model a node's opinion is not only influenced by its neighbors, but also by the node's own prejudices. The relative influence of the prejudices and the neighbors' opinions is controlled by a \emph{stubbornness parameter} that can be different for every node. 
We consider a particular version of the FJ model with only two kinds of nodes (two possible values for the stubbornness parameter): stubborn nodes that only trust their prejudices (\emph{elite}) and nodes with no prejudices (\emph{listeners}). 

Formally, let $G$ be a directed and weighted network with a set $V$ of nodes, a set $E$ of links and an adjacency matrix $A$ with $A_{ij}$ the weight of the link $i\rightarrow j$. We adopt the convention that when there is a link from $i$ to $j$, $i$ is being influenced by $j$ ($j$'s opinion will pass onto $i$ during the inference process)\footnote{In addition to simplifying the notation, this convention also enables a straight forward analysis of retweet networks, which we will use to illustrate this methodology. In a retweet network, when user $i$ retweets user $j$, we make a link $i \rightarrow j$, and since $i$ is broadcasting $j$'s message, she being influenced by $j$.}. Let $S \subset V$ be the set of opinion seeds or \textit{elite} nodes and $L = V \setminus S$, the set of \textit{listeners}. Each elite node is assigned an opinion vector $\vec{x}_s$ that will remain constant throughout the iterative learning process. We call those extreme opinion vectors the \emph{opinion poles}, and they will be defined below. The listener nodes are initially assigned a neutral opinion vector $\vec{x}_l=\vec{0}$ of the corresponding dimension. The opinion of the listeners is inferred by iteratively averaging the opinion of their neighbors, such that at each time step $t$ the opinion of a given listener $\vec{x}_i(t)$ is given by:

\begin{equation}
\vec{x}_i(t) = \frac{\sum_j A^*_{ij} \vec{x}_j(t-1)}{\sum_j A^*_{ij}}
\label{eq:degroot_model}
\end{equation}

With $A^*_{ij}$ the $i,j$ element of a modified adjacency matrix defined in such a way that elite nodes do not change opinion throughout the learning process. This is done by replacing all their outgoing links by a single self-loop (their rows are changed by vectors of $0$s with a $1$ in the diagonal, so we use the Kronecker delta $\delta_{ij}$):

\begin{equation}
A^*_{ij} =
\begin{cases}
A_{ij} \quad &if \quad  $i$ \in L\\
\delta_{ij} \quad &if \quad $i$ \in S
\end{cases}
\end{equation}

This averaging process is iterated until convergence, which is guaranteed (and the solution is unique) as long as every listener is connected to at least one elite node through a \emph{directed path} \cite{parsegovNovelMultidimensionalModels2017}. If $X(t)$ is the matrix whose rows are the opinion vectors of the listeners, such that in a $k$-dimensional opinion space $\vec{x}_i (t) = [X_{i1}(t),X_{i2}(t),\dots,X_{ik}(t)]$, we consider that $X(t)$ has converged when $\lVert X(t) - X(t-1) \rVert_1 < tol$, where $\lVert M \rVert_1 = \sum_{i,j} | m_{ij}|$ is the entrywise $1$-norm of the matrix $M$. We have set the tolerance to $tol=10^{-6}$.

\subsubsection*{Pole vectors and the opinion simplex}

The cornerstone of our multipolar methodology are the elite opinion vectors or \emph{poles}, as they determine the structure of the ideological space. In bipolar settings \cite{moralesMeasuringPoliticalPolarization2015}, 
a scalar value of $x_s=+1$ or $x_s=-1$ is assigned to the nodes of the two opposing elite subsets, placing the poles at a symmetrical position around zero. To generalize this idea for mutipolar systems, we need to
place the poles at a fixed distance from the neutral ideological position (say $\vec{0}$) and in such a way that the distance between any two poles is the same. A geometrical object that fulfills these requirements is the regular simplex\footnote{A regular $k$-simplex is a generalization of the concept of equilateral triangle (which is a regular $2$-simplex) to $k$ dimensions.}. For a system with $n$ poles we build a regular $(n-1)$ simplex and place the poles on its vertices (therefore, $\vec{x}_i \in \mathbb{R}^{n-1}$). In particular, we have chosen a regular simplex centered at the origin with the vertices at distance $u=1$ from the origin. 
We detail the computational process to obtain the coordinates of the pole vectors for an opinion space of arbitrary dimension in Supplementary Section 1. In Figure \ref{fig:diagram_opinion_spaces}A we show the opinion spaces of dimension $1,2$ and $3$. Notice that the distance between poles is different for opinion spaces of different dimension \cite{hajjaDistancesVerticesRegular2017} (see Table \ref{tab:pole_dist_dim}).

\begin{table*}[tbhp]
	\centering
	\caption{\textbf{Coordinates of the poles and distance between them.} Coordinates of the vertices of the opinion simplex ($\vec{v}_0,\vec{v}_1,\dots$) and distance between them for a regular $k$-simplex of different dimensions with vertices placed at distance $u=1$ from its center.}  
	\begin{tabular}{cccccc} 
	\toprule
Number of dimensions & Distance between poles & $\vec{v}_0$ & $\vec{v}_1$ & $\vec{v}_2$ & $\vec{v}_3$\\ \midrule
		$1$ & $2$ & $+1$ & $-1$ \\
		$2$ & $\sqrt{3}$ & $(1,0)$ & $(-\frac{1}{2},\frac{\sqrt{3}}{2})$ & $(-\frac{1}{2},-\frac{\sqrt{3}}{2})$\\
		$3$ & $\sqrt{8/3}$ & $(1,0,0)$ & $(-\frac{1}{3},\frac{\sqrt{8}}{3},0)$ & $(-\frac{1}{3},-\frac{\sqrt{2}}{3},\frac{\sqrt{6}}{3})$ & $(-\frac{1}{3},-\frac{\sqrt{2}}{3},-\frac{\sqrt{6}}{3})$\\
		$k$ & $\sqrt{2 \frac{k+1}{k}}$ \\
		\bottomrule
	\end{tabular}
    \label{tab:pole_dist_dim}
\end{table*}



\begin{figure}[h!]
\centering
\includegraphics[width=\linewidth]{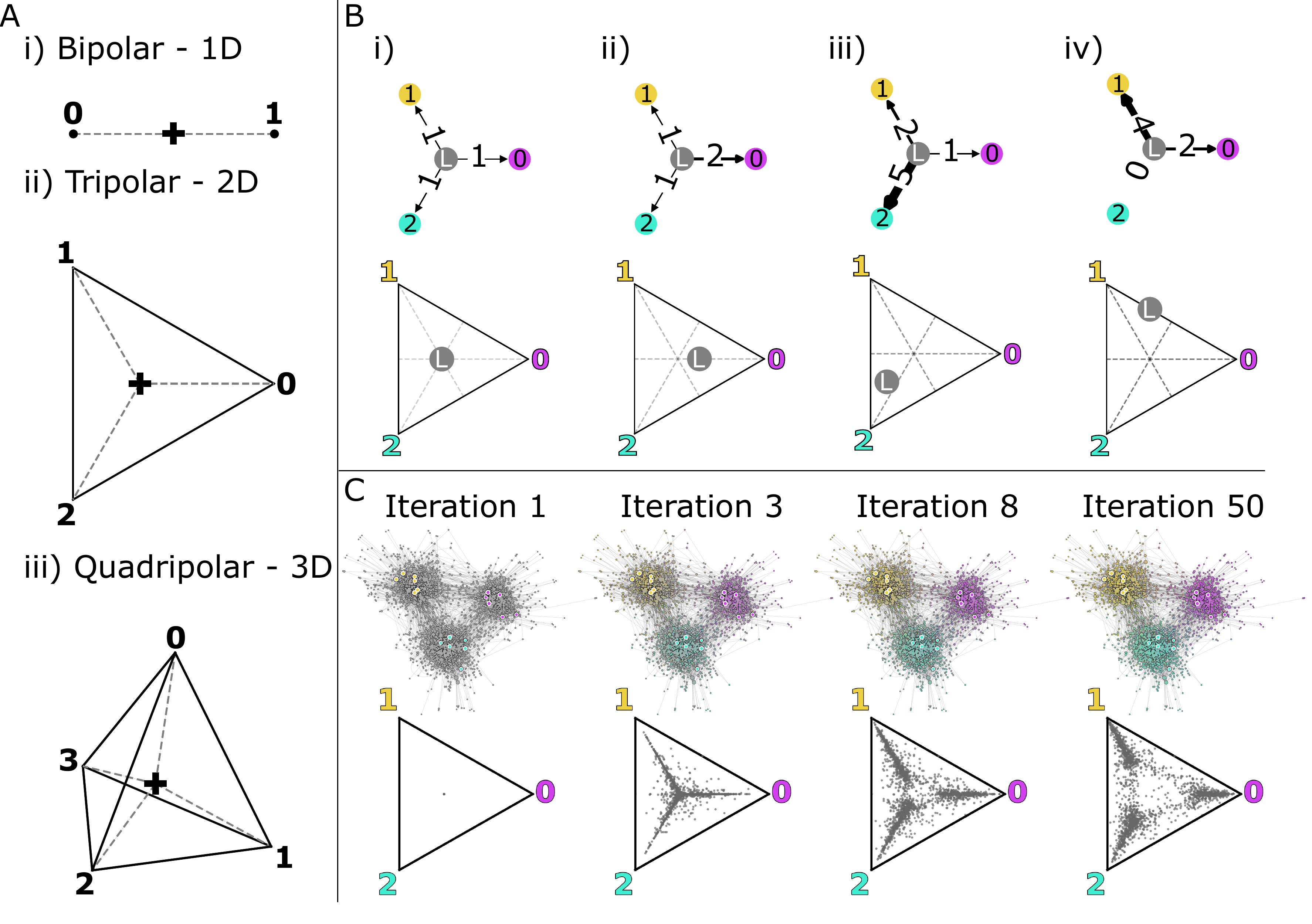}
\caption{\textbf{Illustrating the ideological space and the opinion inference process.} A: Diagram of the opinion spaces of dimension $1$ (i), $2$ (ii) and $3$ (iii). While in 1D the poles can be intuitively labeled as against ($-1$) or in favor ($+1$) of something, in higher dimensions they are labeled with numbers from $0$ to $n-1$ where $n$ is the number of poles. The neutral point (the barycenter of the simplex's vertices) is marked with a black cross. B: Examples of the opinion position of a listener node for three different weighted networks of a tripolar system. Top: network of the tripolar system with one listener node in gray (L) connected to three elite nodes corresponding to the three poles. The numbers on the links are their weights. Bottom: the corresponding opinion simplex showing the opinion position of the listener node. C: Evolution of the opinions of the nodes driven by Eq. \eqref{eq:degroot_model}. The nodes of the networks in the top row are colored according to their closeness to each pole in the ideological space of the bottom row. Interactive versions of panels B and C can be found in \url{https://vis.csh.ac.at/multipolar-viz}.}
\label{fig:diagram_opinion_spaces}
\end{figure}

We set the same distance between every pair of poles to adopt the most neutral, unbiased, or uninformative configuration.
In some contexts it may seem more appropriate to place certain poles closer together; for example, in an election certain political parties may be more ideologically aligned between each other than with the rest. However, parties that share ideology also compete for the same electoral base, often leading to even greater antagonism between them. With our parsimonious approach, we impose no a priori bias and infer the affinity between poles from the opinion distribution obtained from interactions between individuals.

It is worth noting that the opinion assigned to any node will always be located withing the convex hull of the opinion poles, as the opinion updating mechanism is a convex combination and the pole vectors form a regular simplex. Therefore, if a node updates its opinion by increasing its support to a given pole (getting closer to it), her support for the remaining poles will decrease in the same proportion (will get equally away from the others). This implies that each node has the same constant \emph{opinion budget} that it allocates among the different poles (see Supplementary Section 2 for a formal derivation).

In Figure \ref{fig:diagram_opinion_spaces}B we illustrate how the opinion simplex works in a simple tripolar system consisting of a network composed of one listener node connected to three elite nodes, each associated to a different pole. If all the links have the same weight, the opinion of the listener node is neutral (see panel B.i); but if the weight of one of the links increases, the listener's opinion gets closer to the corresponding pole and, importantly, gets \emph{equally away} from the others (see panel B.ii). Panel B.iii shows a situation with different weights for every pole and panel B.iv shows that when the listener is only connected to two of the poles (the remaining weight is 0), its opinion lies on the edge that joins the two poles.

The previous simple example only required one iteration of Eq. \eqref{eq:degroot_model} to obtain the opinion vector of the listener node. In larger networks it takes several iterations to achieve convergence. To better illustrate the process we will use a tripolar synthetic network built with the Lancichinetti-Fortunato-Radichi (LFR) method \cite{lancichinettiBenchmarkGraphsTesting2008}. This method generates scale-free modular networks with a given number of nodes (we have chosen 1300), a given number of communities (we have chosen 3, to simulate a tripolar system), and a certain probability of each node to be connected to nodes of other communities rather than their own. The higher this probability, the more mixed is the network and less distinguishable are the modules. We have set this parameter to $\mu=0.1$. For the opinion inference process, we have built the three subsets of elite nodes of the three poles by ranking the nodes according to their degrees and choosing the top 6 nodes of each community as the opinion seeds of each pole. In Figure \ref{fig:diagram_opinion_spaces}C we show the iterative opinion inference process on this network.




\subsubsection*{Measures of multi-polarization}
\label{sec:multidim_meas_pol}

To study polarization in multipolar contexts we propose several metrics based on the covariance matrix, which is the natural multidimensional extension of the variance, one of the most common measures of social (bi)polarization \cite{DiMaggio1996, Bramson2016, Banisch2019}. 
Firstly, we define a multipolar generalization of the notion of \emph{opinion extremeness}; secondly, we propose a way to compute the directions along which polarization is maximal, allowing us to measure \emph{pole constraint}.

The covariance matrix of a random vector $\vec{Y}$  with expected value $\mathbb{E}[\vec{Y}]$ is defined as follows:

\begin{equation}
Cov[\vec{Y},\vec{Y}] = \mathbb{E}[(\vec{Y} - \mathbb{E}[\vec{Y}])(\vec{Y} - \mathbb{E}[\vec{Y}])^T]
\end{equation}

The trace of the covariance matrix can be interpreted as a measure of \emph{multidimensional variance}, usually called \emph{total variation} \cite{jolliffe2002principal} ($TV$). Moreover, as we show in Methods, when the distance between the barycenter of the opinion simplex and the poles is $u=1$, the maximum attainable $TV$ is $1$, so this metric is normalized by design. 
Crucially, the maximum $TV$ is achieved when there are only extreme opinions and they are uniformly distributed among all the poles. Therefore, the $TV$ can be used as a measure of global polarization combining the aspects of \emph{opinion extremeness} and \emph{community fragmentation} \cite{Bramson2016}; that is, it not only measures how extreme the opinions are but also how evenly is the population divided into the considered factions. 

To characterize and quantify \emph{pole constraint} we compute the eigen-decomposition of the covariance matrix, a technique commonly referred to as 
\emph{Principal Component Analysis} (PCA) \cite{jolliffe2002principal}. The eigenvectors (or Principal Components - PCs) form an orthogonal basis of the opinion space and their corresponding eigenvalues are the variances of the opinion distribution projected onto the direction of their associated eigenvector. Therefore, the eigenvector with the largest eigenvalue (or first Principal Component) corresponds to the direction of maximum variance; that is, the direction of maximum polarization \cite{schweighoferAgentbasedModelMultidimensional2020}. We will show that the projections of the opinion distribution onto the directions of maximum polarization reveal the latent ideological space of the system. The proportion of variance explained by the first PCs provides a measure of \emph{pole constraint}. If it is high, it means that the system's ideological structure has a lower dimensionality than the considered space (it is \emph{constrained} to a subspace). PCA allows us to extract the natural ideological dimensions that organize the social tensions of the system.

\begin{figure}[tbhp]
    \centering
	\includegraphics[width=\linewidth]{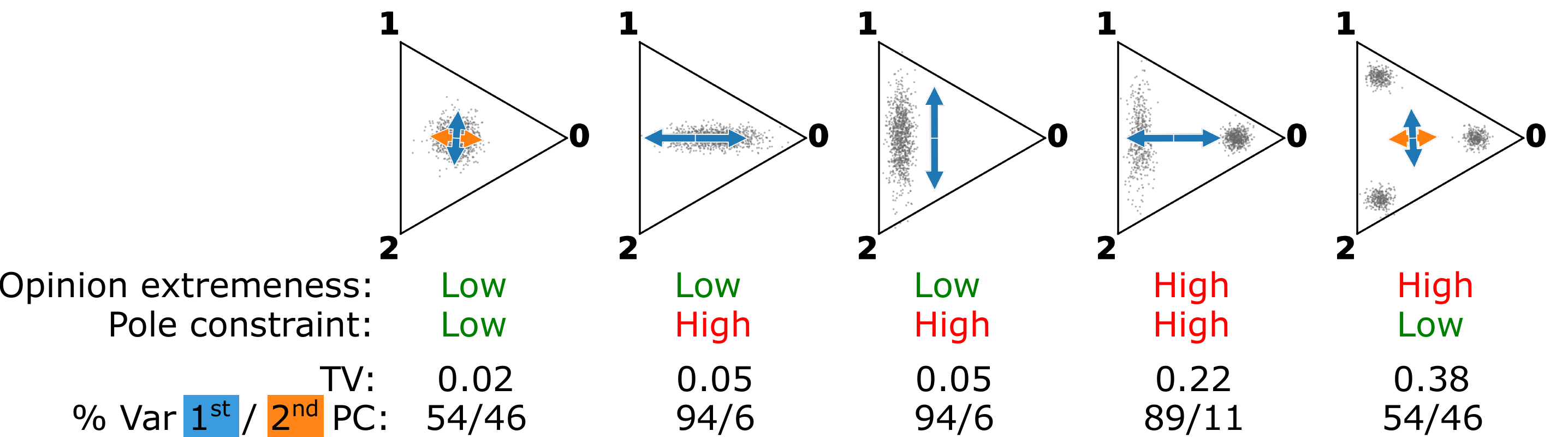}
	\caption{\textbf{Visualization of multi-polarization measures}. Synthetic opinion distributions of tripolar systems and the direction of their corresponding first (blue) and second (orange) PCs shown as arrows with a length proportional to the fraction of the variance they explain. In the three central panels the second PC explains so little variance that is not visible in the plot. The $TV$ of the system (measure of opinion extremeness) and the fraction of explained variance by each PC (measure of pole constraint) is also shown.}
	\label{fig:pca_examples}
\end{figure}

To help build an intuition of these polarization metrics, in Figure \ref{fig:pca_examples} we present synthetic opinion distributions on a tripolar system with different levels of opinion extremeness and pole constraint, respectively measured by their $TV$ and the relative amount of variance explained by the PCs.

\subsection*{Empirical study of multipolar systems}

In this section we illustrate the application of our opinion inference and multi-polarization measurement methodology to two real-world scenarios, one with four poles (3D opinion space) and another with five poles (4D opinion space).  

We analyze these systems using Twitter data. 
We use the retweet networks of the Twitter users who participate in the conversation as the \emph{interaction networks} on which the opinion inference process is performed. Retweets can be considered as a proxy of influence because they are a broadcasting mechanism that usually imply that the retweeting user agrees with the original tweet and has enough interest to perform the retweet action \cite{metaxasWhatRetweetsIndicate2015,yaqubAnalysisPoliticalDiscourse2017,martin-gutierrezRecurrentPatternsUser2018,borondoCharacterizingModelingElectoral2012}. Hence, whenever user $i$ retweets a message originally posted by user $j$, $i$ is being influenced by $j$'s ideas and we make a directed link from $i$ to $j$. The weight of the $i \rightarrow j$ link is the number of times $i$ has retweeted $j$ within a given time interval. 

Here we will focus on the multipolar analysis of the multidimensional opinion distributions, but the preliminary computations to obtain those distributions are detailed in Supplementary Sections 4 and 5, including the selection of the elite and the listeners. 
Throughout the analysis, we will show that the emergent patterns are consistent with the underlying sociopolitical reality and can be interpreted taking into account their context, highlighting the usefulness of the methodology for the study of multipolar social systems. 

\subsubsection*{Quadripolar system: Spanish General Elections of 2015}


We begin by analyzing the Spanish general elections of the 20 of December 2015, with four poles corresponding to the four main political parties. For this study we have worked with Twitter data retrieved during the electoral campaign of 4/12/2015 - 21/12/2015 by searching for messages that include a set of keywords relevant to the context, like \emph{EleccionesGenerales2015} or \emph{EleccionesGenerales28A}. The details of the dataset can be found in Supplementary Section 6, and in Supplementary Section 7 we provide an overview of the sociopolitical context.
The elite of this electoral context are politicians and supporters of PP, PSOE, Podemos and Cs. An elite node is required to have at least $100$ retweets and a participation ratio of at least $50\%$. Additionally, it must belong to one of the communities associated with one of the poles (see Supplementary section 4): 



\begin{figure}[h!]
	\centering
	\includegraphics[width=\linewidth]{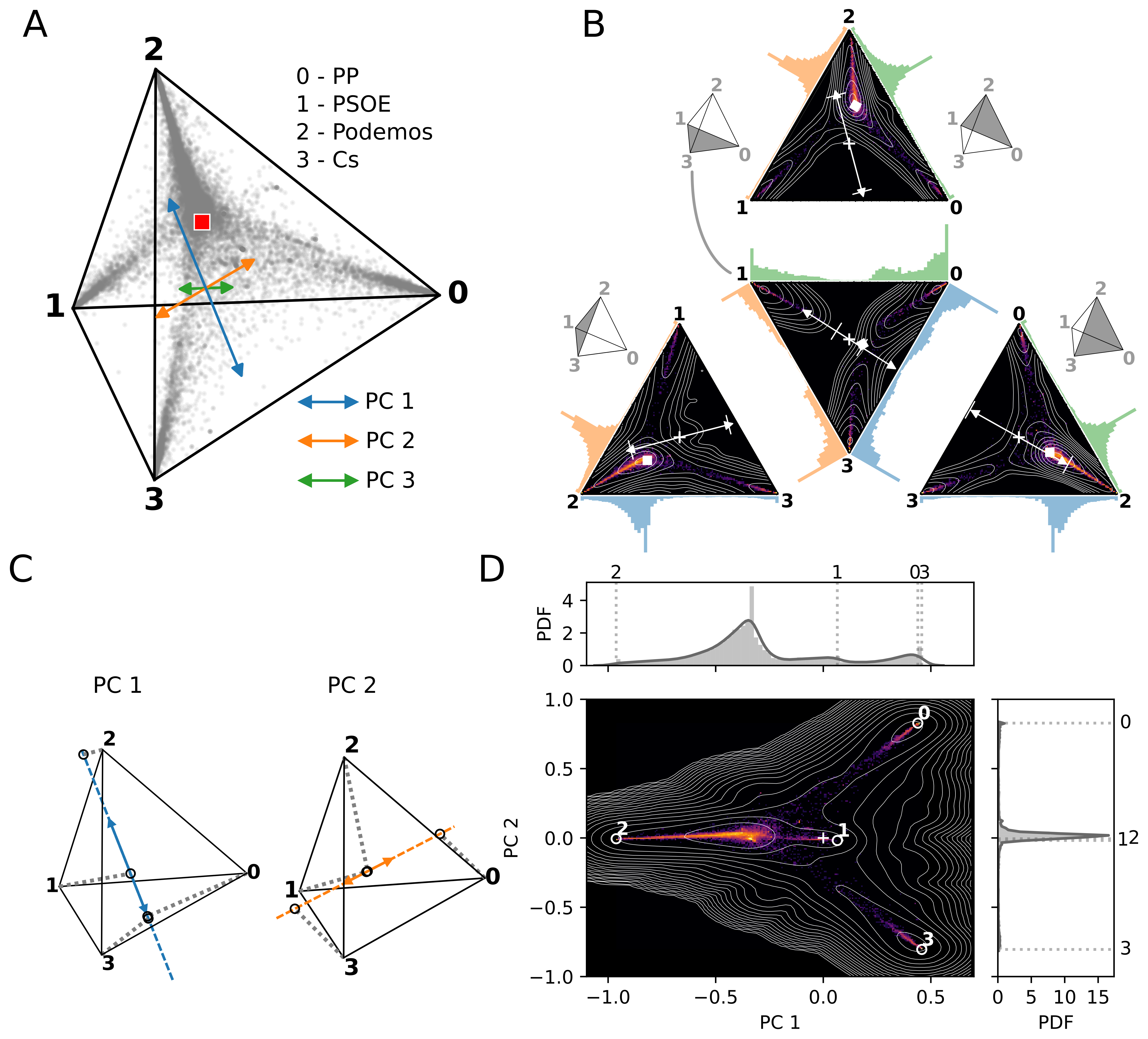}
	\caption{\textbf{Quadripolar opinion distributions of the 2015 Spanish elections.} A: Cloud of users' opinions with its center of mass marked as a red square and the principal components as arrows with length proportional to their respective explained variance. An interactive version of this panel can be found in \url{https://vis.csh.ac.at/multipolar-viz}. B: Projections of the distribution onto the simplex faces shown as heat maps and contour plots. The centers of mass of the projected opinion distributions are represented as white squares and the projection of the direction of maximum polarization (PC 1), as a double headed arrow. The 1D projections onto each edge of the simplex are shown on the sides of the triangles. C: Diagram showing the projection of the poles onto the first two principal components. D: Opinion distribution projected onto PC1 and PC2. Each point has been projected as in the diagram of panel C, which shows the poles' projections.
	}
	\label{fig:2d_od_eg_1516}
\end{figure}

In a quadripolar context, the opinion space is a three-dimensional simplex (a tetrahedron). Since the original 3D opinion distribution is difficult to visualize in detail (see Figure \ref{fig:2d_od_eg_1516}A), we have projected it onto the faces of the simplex, as shown in Figure \ref{fig:2d_od_eg_1516}B. To avoid noise, we have filtered out users with low activity (tweets $<10$). Additionally, in each projection, we only show the opinions of the users that are closer to the poles of interest (the vertices of the face in 2D projections and the vertices of the edge in 1D projections), i.e., opinions within the Voronoi cells of the corresponding poles. As explained in Supplementary Section 3, we do this to provide the most relevant information for each projection. 



Figure \ref{fig:2d_od_eg_1516} shows that users' opinions are mainly concentrated near the poles, with \emph{filaments} emerging from them and stretching towards the point of highest density. This maximum is found halfway between the Podemos pole (2) and the barycenter (marked in the projections with a cross), with the center of mass (marked with a square) very close to that maximum. The location of this maximum also causes the direction of maximum polarization (the first Principal Component, PC 1) to be mostly oriented with Podemos on one side and the other parties on the other.
This can be explained both by the ideology (which will be discussed below) and the popularity of Podemos. In 2015 it was an emergent party that had achieved a moderate success in the previous European elections of 2014 (their first election), and had even reached first place in some electoral polls \cite{Kassam2014, OLeary2015}. Podemos' popularity in Twitter has been also reported in other studies \cite{Guerrero-Sole2018}. Since our multi-polarization measures are based on the variance of the opinion distribution, it is natural that a pole with high support (which concentrates a lot of opinion points in its vicinity) drives polarization.

The white double arrows shown in the projections mark the direction of maximum polarization projected onto each face. PC 1 explains $46\%$ of the total variation, which is as a sign of moderate to high \emph{pole constraint}, meaning that opinions are aligned along a 1D axis to a certain extent. The total variation is $TV=0.23$, 
so this system presents moderate overall polarization in the sense of \emph{opinion extremeness}.

The length of the segment delimited by the two short lines that crosses each white arrow is proportional to the size of the projection of PC 1 in the corresponding face. So when the segment is long, PC 1 is mostly parallel to the face, while when it is short, PC 1 is mostly orthogonal to it. The most orthogonal face to the direction of maximum polarization is $(0,1,3)$, corresponding to PP, PSOE and Cs. This result can be explained because, since Podemos is on the left of PSOE (center-left) and Cs was defined as a center party \cite{Minder2015}, $(0,1,3)$ is the pole triplet with highest ideological affinity. This can also be observed in Supplementary Figure 6A, which shows the ideological location of the parties according to Spanish citizens' perceptions.

Since the principal components reveal the directions along which social tensions are organized, we can also use them to extract the structure of the latent ideological space. In that space, the distances between the parties will no longer be fixed, but determined by their respective affinity according to the opinion distribution. We expose this latent space by projecting the opinion distribution and the positions of the poles onto the PCs. For the sake of visualization we do this for PC1 and PC2, which together explain $78\%$ of the variance. Figure \ref{fig:2d_od_eg_1516}C schematizes the projection process and Figure \ref{fig:2d_od_eg_1516}D shows the resulting distribution.

By looking at the projection along PC 1, we see that the parties are spontaneously divided into left-wing and right-wing. Therefore, even without imposing any such structure on the ideological space a priori, this information is encoded in the opinion distribution. But ideology is not the only driver of antagonism; for instance, parties that support similar policies also compete for the same electoral base, so they tend to emphasize their doctrinal differences \cite{daltonQuantityQualityParty2008,Downs:1957}. This \emph{secondary} source of conflict is manifested in the distance between Podemos and PSOE. These two parties have some ideological affinity: they negotiated to reach a government agreement in 2015 and formed a joint executive after the elections of the 10 of November 2019. However, they also competed to be the leading party on the left, which triggered frictions between them  \cite{Simon2016}. In the case of the right-wing parties, this phenomenon does not manifest itself in PC 1, but it appears very clearly in PC 2, where they are located at opposite extremes.

\subsubsection*{Pentapolar system: Spanish General Elections of 2019}

In this section, we analyze the Spanish General Elections of the 28 of April 2019 as a pentapolar case study. During this election, the support for the four main parties (PP, PSOE, Podemos and Cs) remained high, but a previously minority far-right party called Vox had attracted much attention, causing a transition towards a pentapolar system. In Supplementary Section 7 we give more details about the context in which this transition took place.

\begin{figure}[h!]
	\centering
	\includegraphics[width=\linewidth]{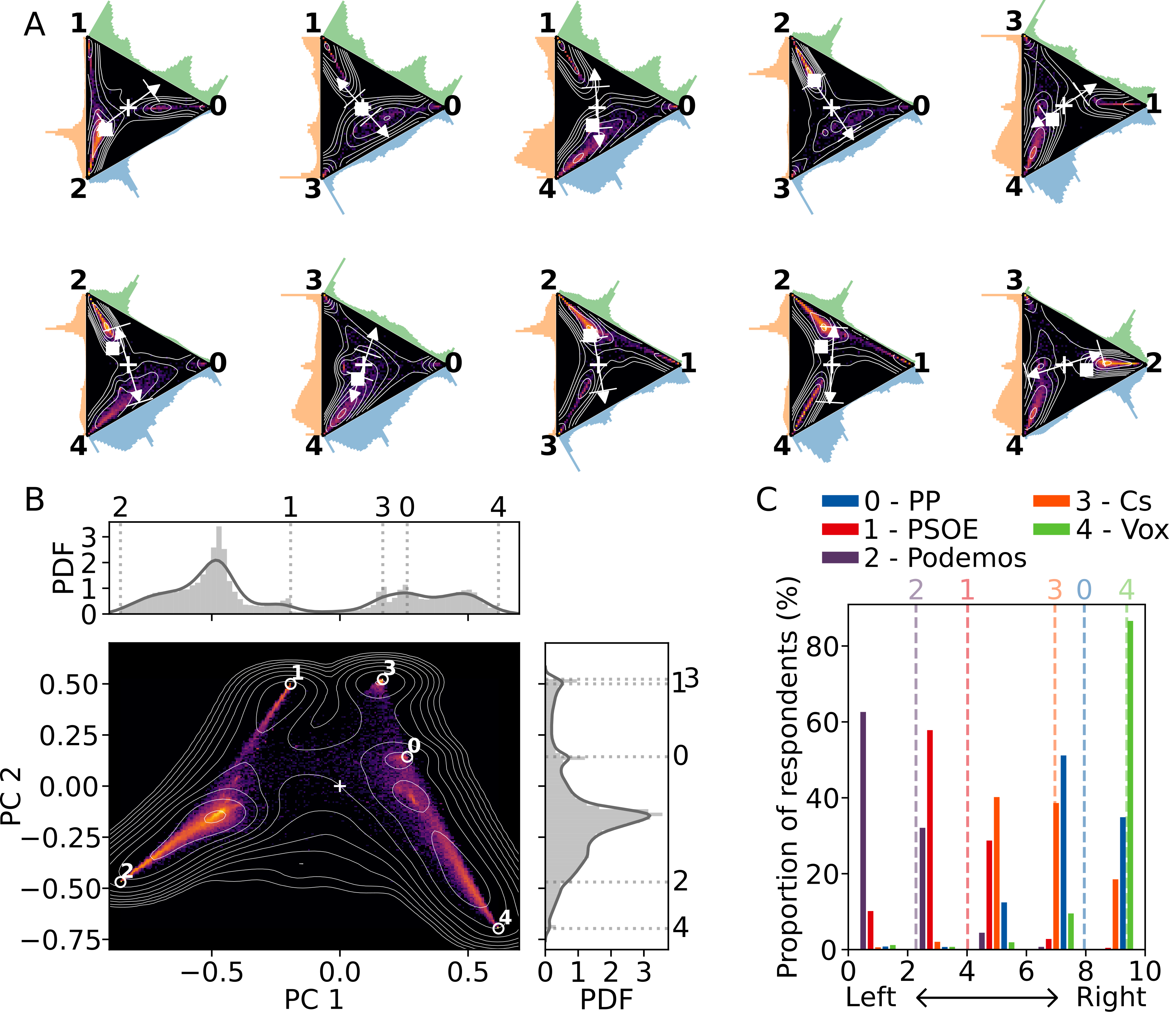}
	\caption{\textbf{Pentapolar opinion distribution of the 2019 Spanish elections.} A: Contour plots corresponding to the 2D projections of the opinion distribution of the 2019 Spanish elections onto the faces of the simplex. The centers of mass of the projected opinion distributions are represented as white squares and the direction of maximum polarization as a double-headed arrow. The 1D projections onto each edge of the simplex are shown on the sides of the triangles. As in Figure \ref{fig:2d_od_eg_1516}, we have filtered out users with low activity (tweets $<30$) to avoid noise, and in each projection we have only considered the opinions of the users that are closer to the poles of interest. B: Projection of the opinion distribution and the poles onto the first two principal components. C: Citizen placement of parties on the left/right scale according to a nationwide survey carried out in May 2019 \cite{cis_data}. The distribution of the answers is shown for every party as a bar plot and the average as a vertical dashed line.}
	\label{fig:2d_od_eg_19}
\end{figure}

For this analysis we have worked with Twitter data retrieved during the period 11/4/2019 - 29/4/2019 (see Supplementary Section 6 for more details). The elite sets for this electoral context have been selected from the communities of influential and engaged users (at least $1000$ retweets and a participation ratio equal or greater than $70\%$) associated to the five relevant parties (see Supplementary section 4):




In a pentapolar system the analysis grows considerably in complexity, as the opinion simplex has 10 faces. The 2D projections of the opinion distribution onto those faces are shown in Figure \ref{fig:2d_od_eg_19}A. 

The highest concentration of users can be found near Podemos and Vox, although Podemos is more popular according to the position of the center of mass (see the projections where poles $2$ or $4$ appear). Additionally, high density areas appear near the edges shared by the two left-wing parties (Podemos and PSOE), as can be appreciated in projections $(0,1,2)$, $(1,2,3)$ and $(1,2,4)$. 
Similar concentrations are also found near the edges shared by two right-wing parties (Cs, PP and Vox) in the projections of 28A, like $(0,2,4)$, $(0,1,4)$,  $(1,3,4)$ or  $(0,1,3)$.

With respect to the behavior of PC 1, faces $(0,1,3)$ and $(0,3,4)$ are the most orthogonal to the direction of maximum polarization, which can be explained by ideological affinity, as poles $(3,0)$-(Cs, PP), $(3,1)$-(Cs, PSOE) and $(0,4)$-(PP, Vox) have formed coalitions in some regional governments. PC 1 explains $57\%$ of the total variance, indicating a high polarization in the sense of pole constraint. Notice that, if the variance was evenly distributed between the four PCs, each would account for $25\%$, and PC 1 is responsible for more than double that value in this case. With a total variation of $TV=0.34$, this system is moderately polarized in the sense of opinion extremeness.

In Figure \ref{fig:2d_od_eg_19}B we show the projection of the opinion distribution onto the first two principal components, which account for $75\%$ of the variance. As in the 2015 elections, in PC 1 the poles are spontaneously divided into left-wing and right-wing parties. And beyond that, their order strikingly matches their ideological positions according to the citizens' perceptions (see Figure \ref{fig:2d_od_eg_19}C). Therefore, in this context the left-wing / right-wing axis completely controls the direction of maximum polarization. The polarization along this axis is so strong that the projected distribution presents a gap (a global density minimum) between the two groups. 
These results are a quantitative validation of the classical notion of left/right ideological axis that has been extensively used to study party systems \cite{daltonQuantityQualityParty2008}.

Once the effects of the left/right tensions have been discounted, in PC 2 the parties get split into extremist and moderate, with poles Podemos (2) and Vox (4) on one side and the rest on the other. Moreover, the relative distances of the poles to the endogenous ideological center matches their order in PC 2. We define the endogenous ideological center as the middle point between the two extreme parties; that is, the intermediate point between Podemos' and Vox's averages in Figure \ref{fig:2d_od_eg_19}C, which is $z_c = 5.83$. The average position ($\bar{z}$) and the distance of each party to the center ($|\bar{z} - z_c|$) are presented in Table \ref{tab:parties_locations_28A}. 

\begin{table}[tbhp]
	\centering
	\caption{\textbf{Perceived level of extremism of political parties.} Average values of the ideological positions of the parties ($\bar{z}$) according to Spanish citizens' perceptions (shown in Figure \ref{fig:2d_od_eg_19}C as dahsed lines). The parties' extremism is measured as the distance of these averages to the endogenous ideological center $z_c$ (middle point between the averages of Vox and Podemos).}  \label{tab:parties_locations_28A}
	\begin{tabular}{cccccc} 
	\toprule
        & Vox (4) & Podemos (2) & PP (0) & PSOE (1) & Cs (3)\\ \midrule
	$\bar{z}$	& 9.38 & 2.28 & 7.95 & 4.02 & 6.96 \\
	$|\bar{z} - z_c|$ & 3.55 & 3.55 & 2.12 & 1.81 & 1.13 \\
		\bottomrule
	\end{tabular}
\end{table}

\section*{Discussion}

In this work we have developed a framework to infer multidimensional opinions and measure different aspects of polarization in multipolar systems. These kinds of systems are common (the most representative example being multi-party democracies), but despite the increasing interest in modeling multidimensional polarization dynamics \cite{doi:10.1073/pnas.2102139118,baumannEmergencePolarizedIdeological2021, schweighoferAgentbasedModelMultidimensional2020,PhysRevE.104.034303}, an analytical framework for characterizing real-world mutlipolar systems was lacking.

One caveat about the datasets used to illustrate the application of this framework is that Twitter data may contain some biases due to the type of audience using it.
Two potential de-biasing strategies could be (i) using complementary data from other online social networks and / or (ii) calibrating the results using surveys. However, we have shown that even without applying any de-biasing, we obtain meaningful results with a clear sociological interpretation.

As for the method itself, it is worth emphasizing that opinion inference and polarization measurement are two independent processes. Therefore, the opinion inference technique can be used on its own in any networked system to obtain a cloud of multidimensional opinion points. This opens the door to the application of any multivariate statistical techniques and machine learning models. In our case studies, for example, clustering techniques like Gaussian mixture models could be used to identify probability density maxima, which can be interepreted as \emph{endogenous} opinion poles. On the other hand, the multipolar analysis based on the covariance matrix is of course not restricted to opinion distributions obtained with our method, but can be used for any opinion distribution \cite{schweighoferAgentbasedModelMultidimensional2020}. The flexibility of our framework makes it suitable to address pressing societal issues.

\paragraph{\textbf{De-escalation of social tensions and working towards consensus.}} As discussed in the Introduction, there is a growing concern about the pernicious influence of polarization on democracy. 
It has been argued that multipolarity could be a possible pathway towards reducing polarization through a more complex information space and greater diversity \cite{doi:10.1073/pnas.2113843118}. But does the diversity of multipolar societies actually alleviate opinion extremeness? Does it help cross-cutting cleavages? As our empirical results show, our analytical framework can significantly contribute to answer these questions.
Beyond their use as descriptive tools, our polarization measures could be used to find the main lines of disagreement in a given context. This could help policymakers devise de-escalation interventions by addressing the issues that cause the strongest tensions. 

\paragraph{\textbf{Mis- and disinformation spreading by bots and trolls.}} It has been argued that the pervasive use of (semi)automatic accounts to spread disinformation is undermining democratic processes by fostering doubt and destabilizing democratic societies \cite{starbirdDisinformationSpreadBots2019}. In some cases, a majority of users, unaware of their role, amplify messages that polarize communities. Our opinion inference technique could help to tackle this issue by evaluating the influence of bots and trolls. Once these accounts have been identified, they can be used as the opinion seeds (the elite) of the inference process. Then, their influence over other users can be compared to that of human actors \cite{gonzalez-bailonBotsAreLess2021}.

From a wider perspective, our results have implications not only for the measurement of polarization, but also for the characterization of party systems.

\paragraph{\textbf{Comprehensive measurement of polarization.}} In line with the literature \cite{Bramson2016,schweighoferWeightedBalanceModel2020}, we argue that polarization is a multi-faceted phenomenon. Its multiple potential meanings complicate its adequate characterization and measurement. Therefore, specifying the notion of polarization under study and its particular operationalization is of utmost importance. Additionally, to fully characterize a system, different measures for complementary aspects of polarization should be used. As we have shown, sometimes it is the less intuitive facets that provide the most relevant information.

\paragraph{\textbf{Characterization of party systems.}} The overlapping collections of policy preferences and ideological identities of parties shape the multipolar opinion distribution. As our analyses of real-world multipolar contexts show, this distribution can be used to infer the structure of the underlying ideological space, including, but not being limited to, the usual left-wing / right-wing axis. This is one of the most commonly used notions to characterize party systems \cite{daltonQuantityQualityParty2008}. Although we have validated the employment of the left/right axis for such analyses, our findings demonstrate that a rigid classification of parties into pre-defined dimensions (left/right, authoritarian/libertarian, etc.) may not paint the most complete or accurate picture. Each context has its own structure, and even within the same democratic system, this structure may and do change. Instead of imposing any structure, our parsimonious approach is flexible and adaptive, finding the most significant dimensions for each context in a principled way.

\section*{Conclusions}


To summarize, we have presented an opinion inference technique, an ideological space model, and a set of multi-polarization measures to characterize multipolar systems. We perform the opinion inference using networks of social interactions. After building the network, we select a small set of elite nodes (opinion leaders) divided into $n$ disjoint subsets, each with a strong affinity for one of the $n$ poles of the system. Then, we use the elite nodes as opinion seeds to estimate the opinion of the remaining nodes (the listeners) with the multidimensional FJ model (an extension of the DeGroot model).

A key contribution of this work is the abstraction of opinions as $(n-1)$ dimensional vectors, with the opinion poles placed at the vertices of an $(n-1)$ regular simplex. The pole vectors defined this way endow the opinion points with the properties of unbiasedness and interdependence. They are unbiased because the poles are placed at a fixed distance from the neutral point and the distance between any two poles is the same. They are interdependent because for an opinion point to get closer to a given pole, it simultaneously must get away from the others.

From the application of the opinion inference technique we get an $(n-1)$ dimensional opinion distribution, which we analyze with multidimensional polarization measures based on the covariance matrix. We use the trace of the covariance matrix as a multidimensional metric for \emph{opinion extremeness} and its eigendecomposition to get the direction of maximum polarization and quantify \emph{pole constraint}.

We have applied this framework to empirical data from multi-party elections, revealing meaningful and complex properties of the systems with direct links to their sociological background. We have found that the ideological differences and affinities between the parties are clearly reflected in the obtained opinion distribution. Those parties with higher affinity tend to form subspaces that are orthogonal to the direction of maximum polarization. Furthermore, the parties' positions along that direction coincide with their location in the left-wing / right-wing axis. However, despite its central role in shaping the opinion space, ideology is not the only driver of polarization. In one of the studied systems, competition between parties of similar ideology was another relevant source of tension. In the other, the perceived extremism of the parties acted as a secondary differentiating axis. Therefore, our approach extracts the drivers of polarization specific to each system. This adaptability makes our framework suitable for the study of democratic systems undergoing political transitions, where the traditional methods based on qualitative sociopolitical analysis may fail to reveal all the relevant information.






\section*{Materials and methods}

\subsection*{Total variation as aggregate multidimensional variance}

To show  that the trace of the covariance matrix (the total variation) can be interpreted as a measure of aggregate multidimensional variance, let us take the definition of variance:

\begin{equation}
Var (Y) = \mathbb{E}[(Y-\mathbb{E}[Y])^2]
\end{equation}

And replace the random variable $Y$ by a random vector $\vec{Y} = (Y_1,Y_2, \dots , Y_k)$, such that the difference between the variable and its expected value is replaced by the Euclidean distance $|| \cdot ||_2$. Then, define the total variation $TV$ as:

\begin{equation}
\begin{aligned}
TV (\vec{Y})& 
= \mathbb{E}[ \lVert \vec{Y} - \mathbb{E}[\vec{Y}] \rVert ^2_2] 
\\
&=\mathbb{E}[\sum_{i=1}^k(Y_i - \mathbb{E}[Y_i])^2] \\
&=\sum_{i=1}^k \mathbb{E} [ (Y_i - \mathbb{E}[Y_i])^2 ] \\
&=\sum_{i=1}^k Var(Y_i) 
= \sum_{i=1}^k Cov[\vec{Y},\vec{Y}]_{ii}
= tr( Cov[\vec{Y},\vec{Y}] )
\end{aligned}
\label{eq:mp_total_var}
\end{equation}

\subsection*{Maximum variance of an opinion distribution}

Opinion distributions are embedded in a regular $k$-simplex with distance $u$ between the barycenter of the simplex and the vertices. This means that the diameter of an opinion distribution, which coincides with the distance between any two vertices of the simplex, is:

\begin{equation}
    d = \sqrt{u^2 \frac{2k+2}{k}}
\end{equation}

Now, let us take into account this theorem proven in a recent work by Lim and McCann \cite{limIsodiametryVarianceRegular2021}, where they use the term variance instead of total variation for the quantity of Eq. \eqref{eq:mp_total_var}:

\begin{quote}
    \textbf{Theorem: Isodiametric variance bound.} If the support of a Borel probability measure $\mu$ on $\mathbf{R}^k$ has diameter no greater than $d$, then $Var(\mu) \leq \frac{k}{2k+2} d^2$. Equality holds if and only if $\mu$ assigns mass $1/(k + 1)$ to each vertex of a regular $k$-simplex having diameter $d$.
\end{quote}

That is, the maximum variance of a probability distribution embedded in a regular $k$-simplex is:

\begin{equation}
    Var_{max} = \frac{k}{2k+2} d^2 = 
    u^2
\end{equation}

So, since we have set $u=1$ for our computations, $Var_{max}=1$. And this value is achieved when there are only extreme opinions (at the vertices of the simplex) and they are uniformly distributed among the poles.

\subsection*{Data and codes}

The interactive visualizations elaborated by Liuhuaying Yang and Johannes Sorger from the visualization team of the Complexity Science Hub Vienna are hosted in \url{https://vis.csh.ac.at/multipolar-viz}. The data and python codes to reproduce the results of the paper are hosted in \url{https://github.com/samuel-mg/multipolar_paper}. The data have been pseudonymized and protected with a code that will be made available by the authors upon reasonable request. We have also prepared easy-to-use python scripts so that researchers can apply the multipolar framework to analyze their own datasets \url{https://github.com/samuel-mg/multipolar}.

\section*{Acknowledgements}

This work has been supported by the Spanish Ministry of Science, Innovation and Universities (MICIU) under Contract No. PGC2018-093854-B-I00, by Spanish Ministry of Education, Culture and Sport (Grant No. FPU15/01461), and by the Austrian research agency (FFG) under project No. 873927. The authors would like to thank Liuhuaying Yang and Johannes Sorger, from the visualization team of the Complexity Science Hub Vienna, for their outstanding visualizations.

\bibliographystyle{ieeetr}
\bibliography{multipole_main_paper}

\end{document}


\maketitle

\section{Cartesian coordinates of the poles}
\label{sec:pole_coord}

To obtain the Cartesian coordinates for our regular $n-1$ simplex we have used the following properties:

\begin{enumerate}[label=\Alph*)]
	\item For a regular simplex, the distance from its center $\vec{0}$ to each vertex $\vec{v}_p$ is the same, and we fix it to $1$:
	
	\begin{equation}
	\left| \vec{v}_p \right| = 1
	\label{eq:sim_ver_norm_1}
	\end{equation}
	
	\item The angle subtended by any two vertices of a $k$-dimensional simplex through its center is $\arccos (\frac{-1}{k})$; therefore, for any two unit vertex vectors $\vec{v}_p$ and $\vec{v}_q$, we have that
	
	\begin{equation}
	\vec{v}_p \cdot \vec{v}_q = \frac{-1}{k}
	\label{eq:sim_ver_ang}
	\end{equation}
\end{enumerate}


The pole vectors $\vec{v}_0, \vec{v}_1, \dots, \vec{v}_{n-1}$ are built by applying properties A) and B) in the following way:

\begin{enumerate}
	\item Set $\vec{v}_0 = (1,0,\dots,0)$.
	\item To compute the first component of $\vec{v}_1$, apply equation \eqref{eq:sim_ver_ang} to vertices $0$ and $1$, which gives us $\vec{v}_1 \cdot \vec{v}_0 = \frac{-1}{n-1}$.
	\item Set all the remaining elements of $\vec{v}_1$ to $0$ except one. 
	\item Compute the only remaining unknown element of $\vec{v}_1$ using equation \eqref{eq:sim_ver_norm_1}.
	\item Compute the first two elements of  $\vec{v}_2$ with equation \eqref{eq:sim_ver_ang}, which gives us $\vec{v}_2 \cdot \vec{v}_0 = \frac{-1}{n-1}$ and $\vec{v}_2 \cdot \vec{v}_1 = \frac{-1}{n-1}$.
	\item Set all the remaining elements of $\vec{v}_2$ to $0$ except one.
	\item Compute the only remaining unknown element of $\vec{v}_2$ using equation \eqref{eq:sim_ver_norm_1}.
	\item Repeat 5-7 for each of the remaining pole vectors.
\end{enumerate}

\section{The opinion budget}

The positions of the poles together with the updating mechanism being a \emph{convex combination} cause that if the support of an individual for a certain pole grows, her support for the remaining poles decreases in the same proportion. This can be easily seen by working in barycentric coordinates, where the vertices of the simplex are the vectors of the standard basis $\vec{e}_j = (0,\dots,1,\dots,0)$. The number of barycentric coordinates is always one more than the number of Cartesian coordinates and the initial position of the listeners (the neutral position) in barycentric coordinates is: $\vec{x}_i(t=0)=(1/n,\dots,1/n)$. The updating mechanism of Eq. (1) of the main document can be written in matrix form as 

\begin{equation}
X(t+1) = W X(t)
\end{equation}

Where $W$ is a matrix whose elements are the normalized rows of the modified adjacency matrix:

\begin{equation}
W_{ij} = \frac{A^*_{ij}}{\sum_j A^*_{ij}}
\end{equation}

Notice that both $W$ and $X(t=0)$ are row stochastic matrices\footnote{The entries of a row stochastic matrix are nonnegative real numbers and the sum of each row is $1$.}, and the product of two row stochastic matrices is row stochastic itself. This means that the \emph{aggregate support} of any listener to all the poles always adds up to one ($\sum_j X_{ij}(t) = 1$ $\forall i \in V$), so if the support for one pole increases, the support for the rest of the poles must decrease accordingly. This notion of \emph{opinion budget} has been studied from the perspective of opinion dynamics by J. Lorenz \cite{Lorenz2008} using multidimensional bounded confidence models.

\section{Choosing the opinions to project}
\label{sec:selec_op_project}

We have considered two different ways to select the listeners whose opinion will be projected: project all of them or project only those that are closer to the projection subspace. 

\begin{figure}
    \centering
	\includegraphics[width=\linewidth]{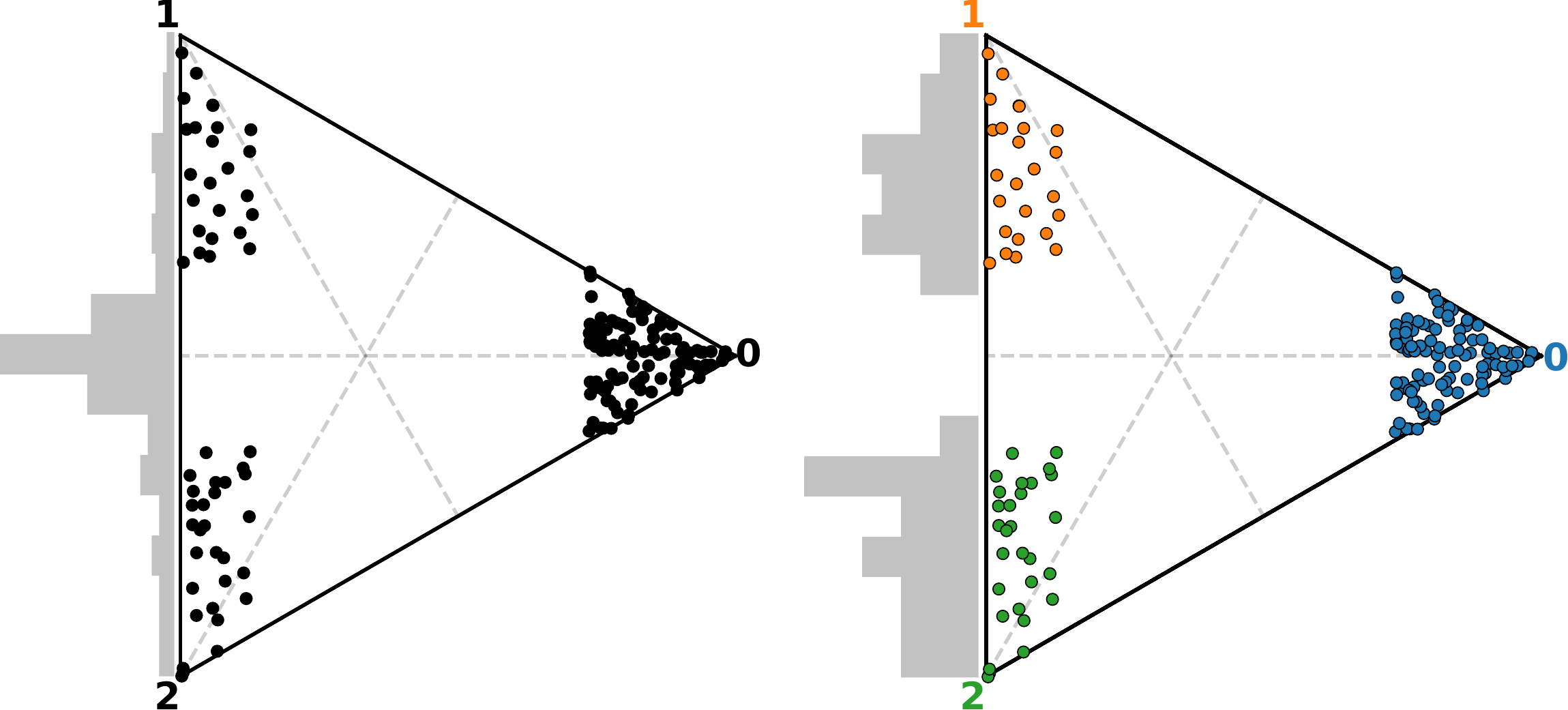}
	\caption{Tripolar opinion distribution projected onto the $(1,2)$ edge when all the listeners' opinions are equally taken into account (left) and when only the opinions of the ones closest to poles $1$ and $2$ are considered (right). The resulting $1D$ distribution is represented as a gray histogram with its maximum rescaled to have a fixed height to facilitate comparison between projections.}
	\label{fig:all_v_close_only_exmpl}
\end{figure}

The most straight forward way is to project the opinions of all the listeners, getting a broad idea of their preference when the options are restricted to the poles that define the projection subspace. However, in certain cases it may not be convenient to project all the listeners' opinions, as the projection may lead us to biased conclusions or misinterpretations. Consider the situation of the left panel of Figure \ref{fig:all_v_close_only_exmpl}, where we have a tripolar system with opinions concentrated near the poles. If we project all the opinions onto the $(1,2)$ edge, the opinion distribution seems to have a peak around the neutral point, which may induce us to conclude that there is a reasonably high consensus in a moderate position regarding poles $1$ and $2$. But looking at the $2D$ distribution we see that the apparent consensus comes from listeners' opinions near pole $0$, which obscure the contributions of the listeners closer to the poles of interest, so our interpretation of the projection at first sight may be wrong. Moreover, it is reasonable to assume that listeners very close to pole $0$ do not care much about the rest of the poles, so their opinions may not be that relevant for this particular subspace.

\begin{figure}
	\centering
	\includegraphics[width=\linewidth]{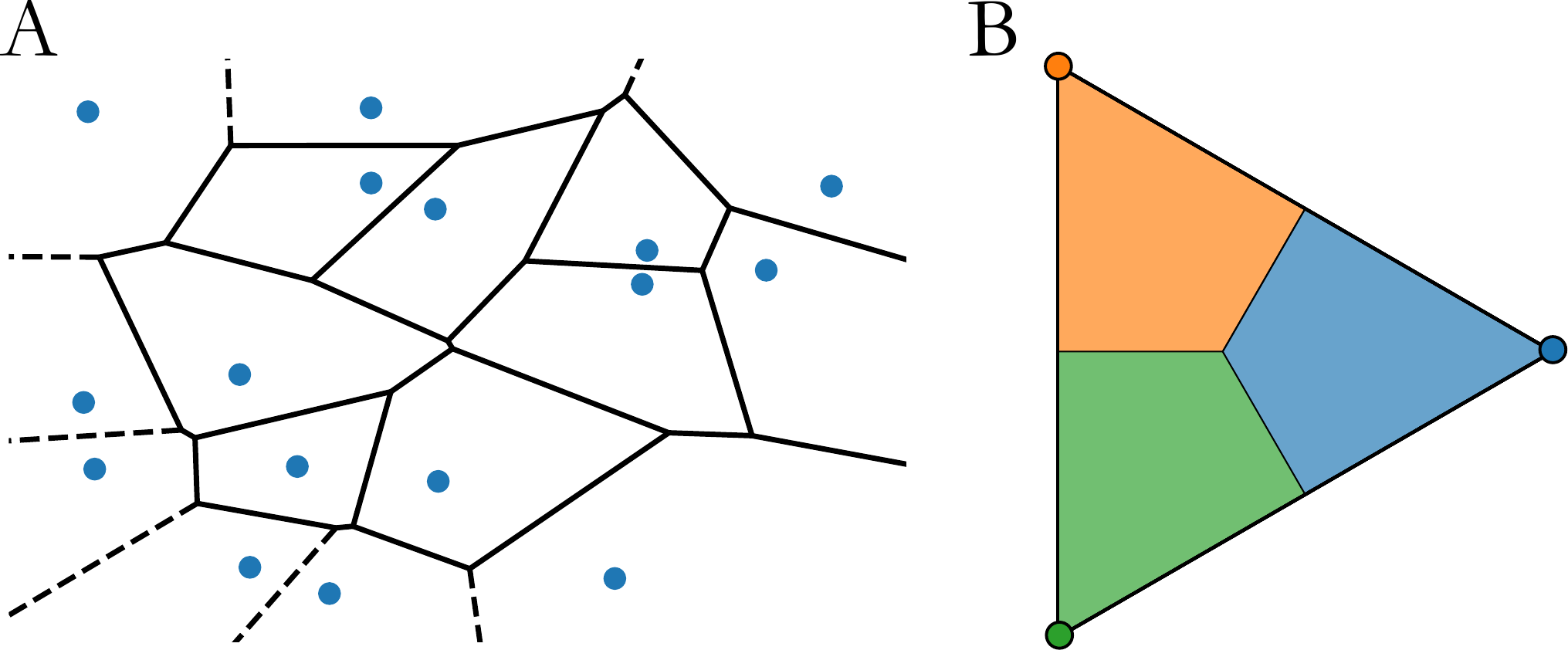}
	\caption{A: Voronoi cells of a cloud of points in the Euclidean plane. Dashed lines are those with an end point in infinity. B: Voronoi cells of the vertices (the poles) of an equilateral triangle (a tripolar opinion space).}
	\label{fig:voronoi_example}
\end{figure}

The second method circumvents these issues by projecting only the opinions of the listeners that are closer to the poles of interest. When we do that we obtain the right panel of Figure \ref{fig:all_v_close_only_exmpl}, where the bimodal nature of the opinion distribution in the $(1,2)$ axis is clearly exposed. The selected opinion points in this case are those contained in the Voronoi cells of the poles that define the subspace. Given a set of points in an Euclidean space of arbitrary dimension, the Voronoi cell of a given point is the region of the space closer to that point than to any other of the points of the set \cite{10.1145/116873.116880}. Panel A of Figure \ref{fig:voronoi_example} shows the Voronoi cells associated to a set of points in the Euclidean plane. In most of our analysis we project only the opinion points within the Voronoi cells associated to the poles defining the projection subspace. In Panel B of Figure \ref{fig:voronoi_example} we show the Voronoi cells of the poles of a tripolar opinion space.


\section{Elite selection}
\label{sec:elite_selection}

One of the main challenges of applying the multipolar methodology is the selection of the set of nodes that are part of the elite. Elite nodes are crucial since they are the seeds that will influence the opinion of the remaining nodes. Therefore, these nodes should be very influential and should have constant and extreme opinions. This way, the full range of opinions can be inferred. Since we apply the methodology to Twitter data, the elite is chosen by analyzing behavioral patterns of the users (which are the nodes). The selection is carried out in two steps.

First, we look for highly engaged and influential users with a well defined opinion. The engagement of a user is measured as the proportion of days (out of the total) that the user participates in the conversation (the participation ratio). The influence of a user is measured as the number of retweets received. To get a reasonably sized elite set, the specific thresholds for participation ratio and retweets are adjusted to each dataset. Nevertheless, the resulting opinion distribution is robust with respect to fairly large changes in the thresholds \cite{atienza-barthelemyRelationshipIdeologyLanguage2019}. 

Once these highly engaged and influential users are selected, in the second step we check which of them hold extreme and mutually exclusive opinions to assign them to the corresponding opinion pole. We do that by analyzing the whole retweet network with the Nested Stochastic Block Model \cite{PhysRevX.4.011047}. This model is hierarchical, which means that it reveals the community structure of the network at different levels. The first level corresponds to the individual nodes, and the higher the level, the lower the number of communities it has. Several communities of a given level may merge into a larger community in a higher level. The highest level has only one community that corresponds to the whole network.


We calculate the community structure of the retweet network with the aforementioned model and analyze the community assignment of the elite users selected in the first step. Then, we look for the highest level of the community structure where we can identify, in each community, nodes with a known affiliation to only one of the poles. If there are nodes from two or more different poles, we go to an inferior level of the community structure. To identify the factions contextual information is required. For example, in many social issues political parties usually endorse different ideological positions, so a straight forward approach is to look for elite nodes corresponding to political user accounts whose party has explicitly shown support for a given option. Since we have mostly worked with political contexts, this is the method we have followed. The process can be considered as a \emph{snowball} sampling technique, as the user subsets of each pole is built from a small sample of nodes with known affiliation.

In Table \ref{tab:elite_net_general} we present the minimum number of retweets ($RT_{min}$) and minimum percentage of participation days ($PR_{min}$) considered to filter influential and engaged users, as well as the number of users obtained by applying this filter and the number of users finally selected as elite after analyzing the community assignment of the filtered users. The number of listeners and links of the corresponding retweet network is also shown.

\begin{table}[h!]
    \centering
    \caption{Minimum retweets ($RT_{min}$) and minimum participation ration ($PR_{min}$) used to filter the influential and engaged users ({Initial Elite}) for each Twitter conversation. The {Selected Elite} corresponds to the number of users of the {Initial Elite} that have been assigned to a pole taking into account the community structure and contextual information. The number of {Listeners}  and the number of {Links} of the final elite-connected network are also shown.}
	\begin{tabular}{lccccccc}
		\toprule
		System & {Poles} & $RT_{min}$/Tweets & $PR_{min}$/\% & {Initial Elite} & {Selected Elite} & {Listeners} & {Links}
		\\ \midrule
		Elections 2015 & 4 & 100 & 50 & 967 & 342  & 236855 & 730802
		\\
		Elections 2019 28A & 5 & 1000 & 70 & 1485 & 631 & 837056 & 7148478
		\\ \bottomrule
	\end{tabular}
	\label{tab:elite_net_general}
\end{table}

\begin{figure}
	\centering
	\includegraphics[width=\linewidth]{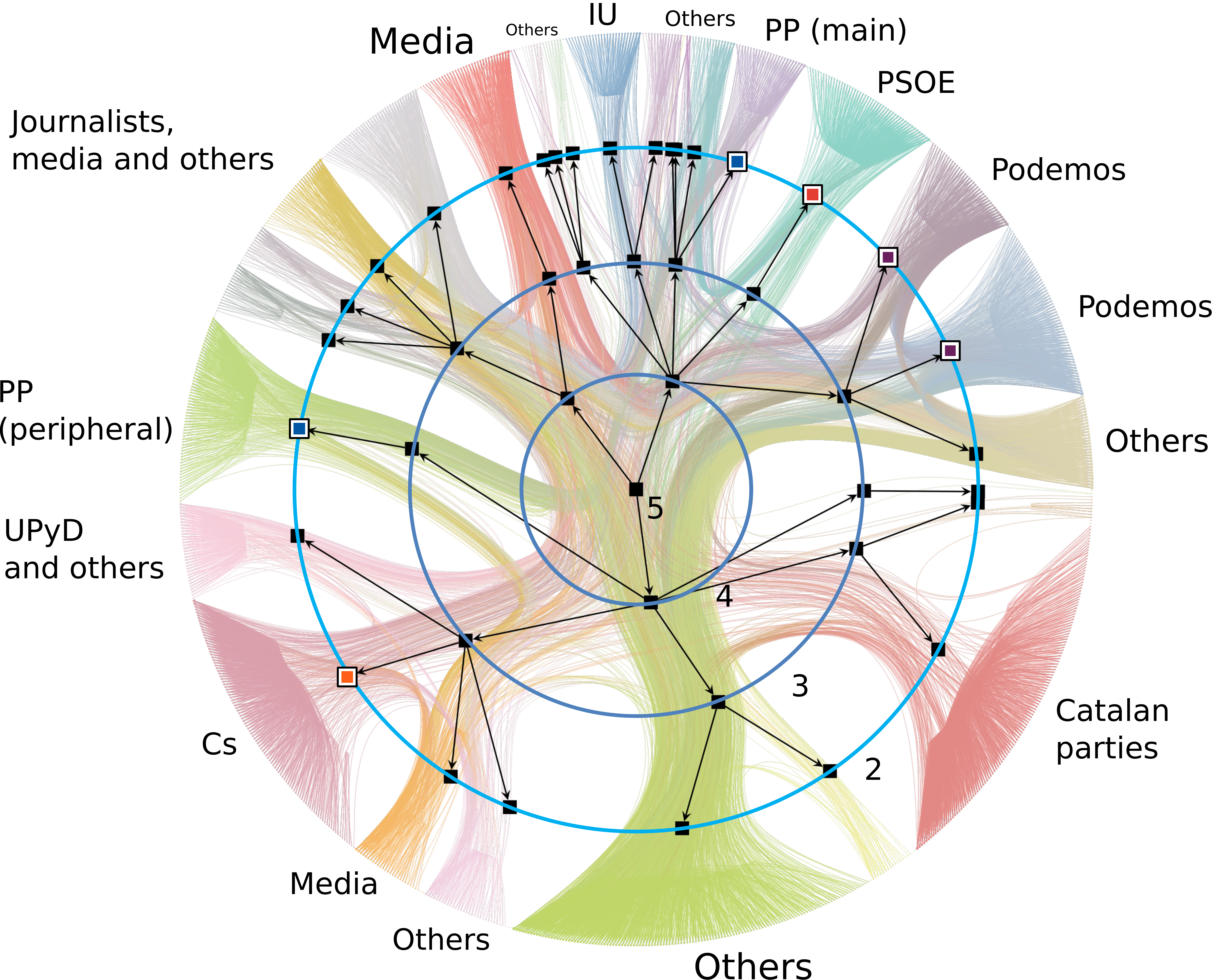}
	\caption{Levels 2 to 5 of the hierarchical community structure for influential and engaged users of the 2015 elections system.  The level used to build the elite is highlighted in light blue and the specific communities assigned to each pole are marked with squares of different colors. Blue: PP. Red: PSOE. Purple: Podemos. Orange: Cs.}
	\label{fig:elite_comm_2015}
\end{figure}

\begin{figure}
	\centering
	\includegraphics[width=\linewidth]{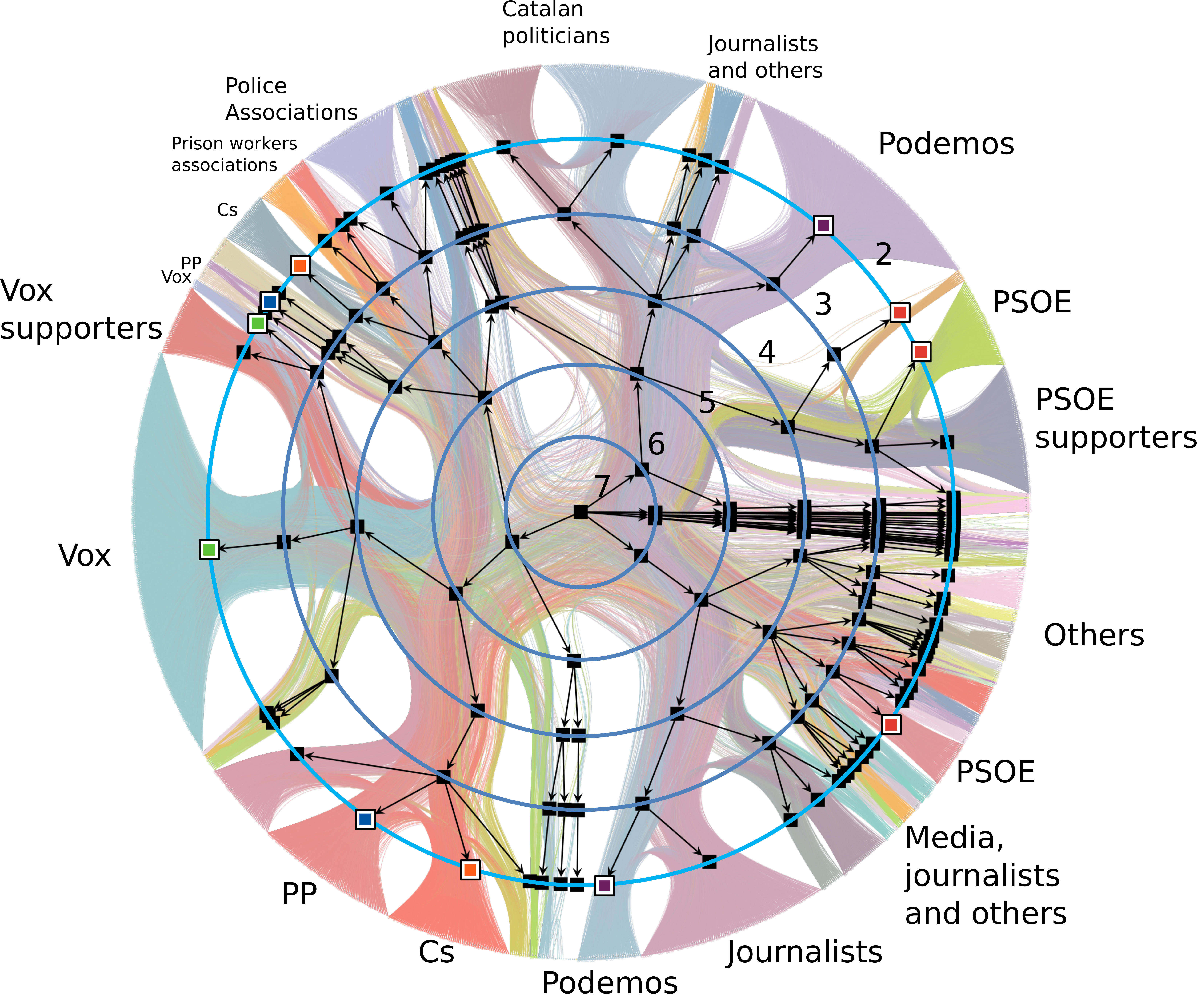}
	\caption{Levels 2 to 7 of the hierarchical community structure for influential and engaged users of the 28A elections system.  The level used to build the elite is highlighted in light blue and the specific communities assigned to each pole are marked with squares of different colors. Blue: PP. Red: PSOE. Purple: Podemos. Orange: Cs. Green: Vox.}
	\label{fig:elite_comm_28A}
\end{figure}

In Figures \ref{fig:elite_comm_2015} and \ref{fig:elite_comm_28A} we show the hierarchical community structure of the influential and engaged users of each system according to the block model. In this hierarchical structure, the first level corresponds to the individual nodes; in the second level, nodes are aggregated in small communities, and the upper levels progressively merge the nodes into larger and larger communities until the top level is reached, where the whole network is in one community. In each figure, the nodes of the outermost circle are the influential and engaged users. The nodes are grouped and colored according to the communities of a certain level. Those same communities are represented as black squares in the second outermost circle. Then, in each inner circle some of the communities of the immediate lower (outer) level merge until all of them are in the same community at the top level. We have labeled the levels such that level 0 corresponds to the one with smallest communities (the level immediately above individual nodes) and the level with highest number corresponds to the full network.

In each system, we have looked for the highest level where the communities contain users from at most one pole (some may contain users with no clear affiliation, like regular users, media, journalists...). In that level we have chosen communities where we can find users with a clear affiliation to a pole and assigned all the users inside those communities to the same pole. To avoid wrongly including neutral nodes in a pole's elite, we have elaborated a list of media accounts (TV, radio, press...) which are removed from every elite set. Even if many media clearly lean towards certain poles, we prefer not to fix their opinions a priori and assign them one with the inference technique to minimize biases.

In Figure \ref{fig:elite_comm_2015} we show levels 2 to 5 of the Spanish elections of 2015. The complexity of this quadripolar system makes it necessary to look for \emph{monopolar communities} at level 2 of the hierarchy and choose more than one community for Podemos and PP poles. As can be appreciated in the figure, the two communities of Podemos merge in the immediate upper level, but the two communities of PP do not merge until the top level. This counter-intuitive phenomenon may arise for two reasons. One reason is related to the properties of the system, as the nodes inside each of the two PP communities belong to different \emph{sections} of the party: in one of the communities we find the main user accounts (the official account of the party, the party's candidate, etc.) and in the other one we mainly find accounts associated to politicians and party branches from peripheral regions of Spain (Andalusia, Galicia and Catalonia for example). The other plausible reason stems from the criterion of the employed block model to assign nodes to one community or another. While many community detection algorithms aim to find groups of nodes with dense intra-connectivity between them and sparse inter-connectivity to the rest of the network, the block model considers all possible connection patterns when grouping nodes in a community. For example, in a star-like network it may place all periphery nodes in one community and all the core nodes in another, as those two blocks provide a very good description of the network topology. Therefore, the separation in two communities may also emerge due to the different communication patterns of the users inside them. In any case, notice that a thick bundle of links joins the two communities. The number of elite users associated to each pole are presented in Table \ref{tab:2015_elite}.


\begin{table}[htbp]
	\centering
	\caption{Number of elite users assigned to each pole of the Spanish elections of 2015.}
	\begin{tabular}{lc}
		\toprule
		{Pole} & {Number of users}
		\\ \midrule
		PP & 89
		\\
		PSOE & 49
		\\
		Podemos & 102
		\\
		Cs & 102
		\\ \bottomrule
	\end{tabular}
	\label{tab:2015_elite}
\end{table}

In Figure \ref{fig:elite_comm_28A} we show levels 2 to 7 of the 28A elections of 2019. In this pentapolar system the analysis grows in complexity and at level 2, the highest with \emph{monopolar communities}, a high community fragmentation can be appreciated. The number of elite users associated to each pole are presented in Table \ref{tab:28A_elite}.


\begin{table}[htbp]
	\centering
	\caption{Number of elite users assigned to each pole of the Spanish elections of 2019 28A.}
	\begin{tabular}{lc}
		\toprule
		Pole & Number of users
		\\ \midrule
		PP & 79
		\\
		PSOE & 77
		\\
		Podemos & 174
		\\
		Cs & 91
		\\
		Vox & 210
		\\ \bottomrule
	\end{tabular}
	\label{tab:28A_elite}
\end{table}

\section{Listeners selection}

\begin{figure}
    \centering
	\includegraphics[width=0.7\linewidth]{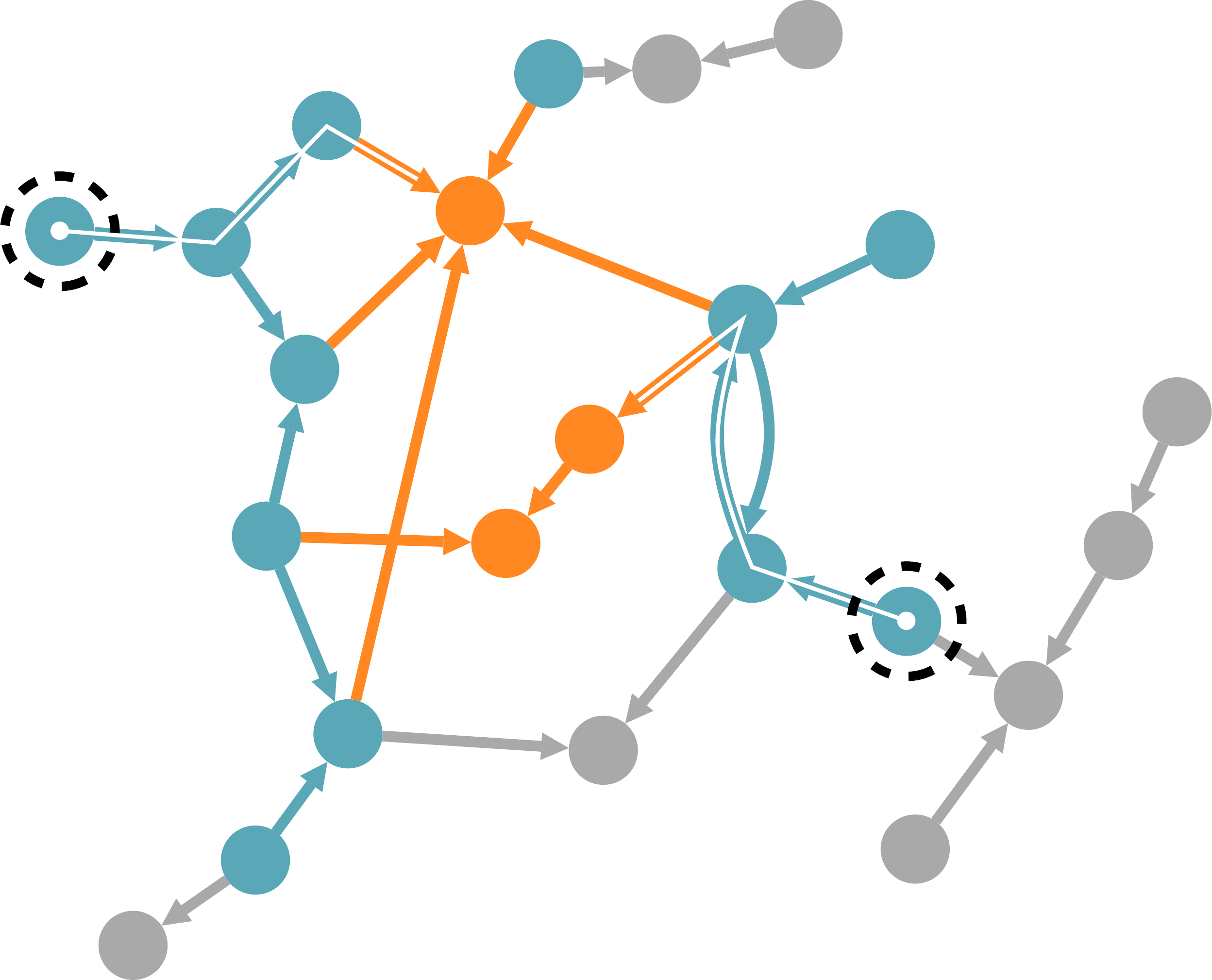}
	\caption{Diagram of a network showing the criterion to select the listeners once the elite has been chosen. Listener users (blue) are those which have a directed path to an elite user (orange). The gray users are those that are neither elite, nor have a directed path to any elite user. Two examples of directed paths from two circled listeners to two elite users are marked in white. Links are colored according to their target node.
	}
	\label{fig:diagram_elite_con}
\end{figure}

After building the retweet network and choosing the elite nodes, we identify the set of listeners, which are all the nodes that are connected to at least one elite node through a \emph{directed path}. This means that there must be a directed path that starts in each listener and ends in at least one elite user. This condition, which is schematized in Figure \ref{fig:diagram_elite_con}, guarantees the convergence of the listener's opinions to a unique solution \cite{parsegovNovelMultidimensionalModels2017}.


\section{Twitter datasets}
\label{sec:datasets_twitter}

We have worked with datasets extracted from Twitter conversations. Each dataset has been retrieved by searching for tweets containing at least one of a given set of keywords during a specific period of time. In table \ref{tab:twitter_datasets} we present the keywords used to build each dataset as well as the number of users that took part in each conversation, the number of tweets they published and the time period considered in each case.

\begin{table}[htbp]
	\centering
	\caption{
		Description of the Twitter datasets. (*) Due to the high quantity of keywords used for the Spanish elections of 2019, we have put them in an online repository.}
    \tiny
	\begin{tabular}{llrrcc}
		\toprule
		\textbf{Dataset} & \textbf{Keywords} & \textbf{No. Tweets} & \textbf{No. Users} & \textbf{Time interval} & \textbf{Time span/days} \\ \midrule
		Spanish elecions 2015 & 20D, 20D2015, \#EleccionesGenerales2015 & 1796093 & 409410 & 2015/12/04 – 2015/12/21 & 18 \\ 
		2019 28A Spanish elections & \url{https://pastebin.pl/view/84efeb1b} (*) & 17706517 & 1339514 & 2019/04/11 – 2019/04/29 & 19 \\  \bottomrule
	\end{tabular}
	\label{tab:twitter_datasets}
\end{table}

\section{Sociopolitical context}


Spain has a bicameral parliamentary system, where the lower house is called Congress of Deputies and the upper house, the Senate. For elections to the Congress of Deputies, held every four years, each of the 50 provinces serves as an electoral district, with the number of deputies representing it determined by its population. Under a proportional representation electoral system governed by the d’Hondt formula, ballots are cast for a provincewide party list rather than for candidates representing individual constituencies.

About four-fifths of the members of the Senate are directly elected via a plurality system at the provincial level. Each province is entitled to four representatives; voters cast ballots for three candidates, and those with the most votes are elected. The remainder of the senators are appointed by the regional legislatures. Because representation is not based upon population, in the Senate smaller and more-rural provinces generally are overrepresented in relation to their overall population \cite{britannica_elections}. Hence, with the end of illustrating the Spanish political landscape, in this work we have chosen to report only the results for the Congress of Deputies, as the distribution of seats is supposed to be more representative.

On 20 December 2015, the Spanish general elections were held. PP (Partido Popular - People's Party) and PSOE (Partido Socialista Obrero Espa\~nol - Spanish Socialist Workers' Party), which constituted the traditional two-party system had lost a lot of social support, while the emerging parties Podemos (We can) and Cs (Ciudadanos - Citizens) were on the rise. This caused a transition from a two-party system to a multi-party system \cite{doi:10.1080/13608746.2016.1198454,FernandezGracia2018}.

In spite of that, PP, which was holding the government, was still leading the polls. The rest of the parties were behind but not too far away. In fact, the supports for the other three main parties fluctuated so much during the year before the election that it seemed impossible to predict, by looking at the polls, which would be the final ranking of votes \cite{wiki_polls15}.

The result was a fragmented parliament where no party held an absolute majority and large coalitions were needed to form a government. The votes and seats that each party obtained are displayed in table \ref{tab_res_elec_2015}. After several months of negotiations there was no agreement between any group of parties large enough to obtain a parliamentary majority that would allow the formation of a government. This led to the announcement of a new election on the 26 June 2016.

\begin{table}[htbp]
    \centering
    \caption{Results of the Spanish general elections of 20 December 2015 and 26 June 2016. Podemos and IU are together in 2016 because they formed a coalition called UP.}
	\begin{tabular}{lcccc}
		\toprule
		& \multicolumn{ 2}{c}{Votes} & \multicolumn{ 2}{c}{Seats} \\
		& $2015$ & $2016$ & $2015$ & $2016$ \\ \midrule
		PP & $7236965$ & $7941236$ & $123$ & $137$ \\ 
		PSOE & $5545315$ & $5443846$ & $90$ & $85$ \\ 
		Podemos & $5212711$ & \multirow{2}{*}{$5087538$} & $69$ &\multirow{2}{*}{$71$} \\
		IU & $926783$ & \multicolumn{ 1}{c}{} & $2$ & \multicolumn{ 1}{c}{} \\ 
		Cs & $3514528$ & $3141570$ & $40$ & $32$ \\ 
		Others & $2775011$ & $2665069$&$26$&$25$ \\ \bottomrule
	\end{tabular}
	\label{tab_res_elec_2015}
\end{table}

Before this new election, one of the emerging parties, Podemos, formed a coalition with IU (Izquierda Unida - United Left). This alliance was called UP (Unidos Podemos - United We Can). Note that UP is simply called ``Podemos'' in the main document.

The 2016 election resulted in a parliament that was almost as fragmented as the one in 2015. The votes and seats obtained by each party are also presented in table \ref{tab_res_elec_2015}. In October 2016, PP finally won the appointment vote and formed a minority government. However, in June 2018, a motion of no-confidence was issued against the ruling party (PP) and PSOE formed a minority government.

Due to the government's inability to pass the national budget bill, another election was held on 28 April 2019. The support for the four main parties (PP, PSOE, Podemos and Cs) remained high, but a previously minority far-right party called Vox had attracted much attention after obtaining parliamentary representation for the first time. In December 2018 they had won 12 of the 109 seats in the regional parliament of Andalusia, and in the 28A general election they gained 24 of the 350 seats in the Congress of Deputies \cite{Hedgecoe2019}. After a failure to form a government due to lack of political agreements, another general election was held on 10 November 2019, when the presence of Vox in the parliament raised to 52 seats. Following the November 10 election, a coalition of PSOE and UP formed a minority government. The votes and seats obtained by each party in the two elections of 2019 are shown in table \ref{tab_res_elec_2019}.

\begin{table}[htbp]
    \centering
    \caption{Results of the Spanish general elections of 28 April 2019 and 10 November 2019.}
	\begin{tabular}{lcccc}
		\toprule
		& \multicolumn{ 2}{c}{Votes} & \multicolumn{ 2}{c}{Seats} \\
		& $2019$ 28A & $2019$ 10N & $2019$ 28A & $2019$ 10N  \\ \midrule
		PP & $4373653$ & $5047040$ & $66$ & $89$ \\ 
		PSOE & $7513142$ & $6792199$ & $123$ & $120$ \\ 
		Podemos (UP) & $2381960$ & $5087538$ & $33$ & $26$ \\
		Cs & $4155665$ & $1650318$ & $57$ & $10$ \\ 
		Vox & $2688092$ & $3656979$ & $24$ & $52$ \\ 
		Others & $4850169$ & $4833311$&$47$&$53$ \\ \bottomrule
	\end{tabular}
	\label{tab_res_elec_2019}
\end{table}

\section{PCA of the opinion distributions}

The eigenvalues of each principal component for the two datasets under study are shown in Table \ref{tab:pca_eigenval}.


\begin{table}[htbp]
    \centering
    \caption{Eigenvalues of the covariance matrix (projected variance) and fraction of the total variance explained by each principal component.}
	\begin{tabular}{lcccc}
		\toprule
		& \multicolumn{ 2}{c}{December $2015$ } & \multicolumn{ 2}{c}{April $2019$} \\
		Principal Component & Eigenvalue & Explained $Var$ (\%)&  Eigenvalue & Explained $Var$ (\%) \\ \midrule
		PC 1 & $0.105$ & $46$ & $0.194$ & $57$ \\ 
		PC 2 & $0.072$ & $32$ & $0.062$ & $18$ \\ 
		PC 3 & $0.052$ & $23$ & $0.049$ & $14$ \\
		PC 4 & $-$ & $-$ & $0.036$ & $11$ \\ 
	\end{tabular}
	\label{tab:pca_eigenval}
\end{table}

\section{Ideological location of the parties according to citizens' perceptions}

\begin{figure}
    \centering
	\includegraphics[width=0.8\linewidth]{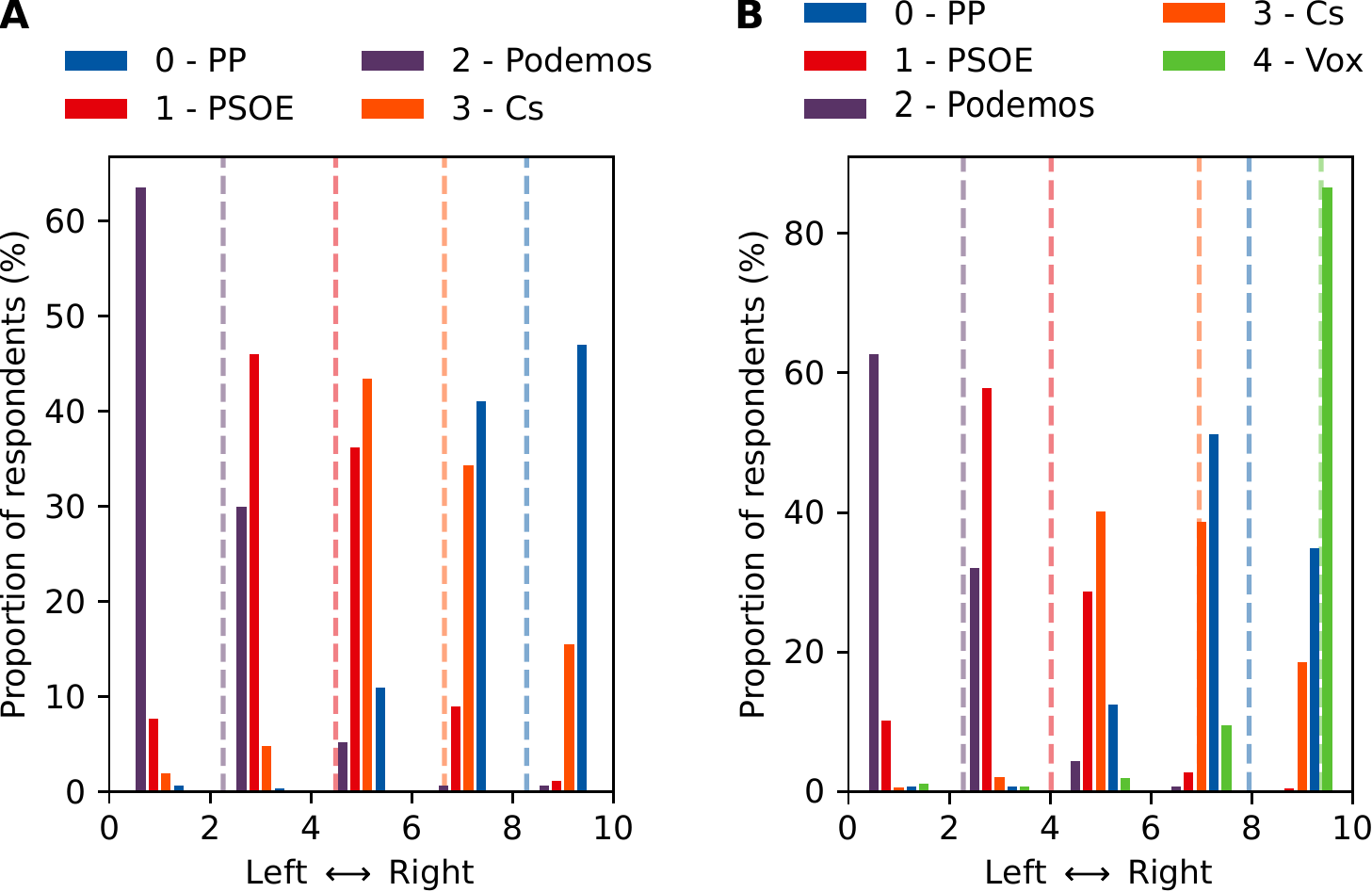}
	\caption{Citizen placement of parties on the left/right scale according to a nationwide survey\cite{cis_data}. The distribution of the answers is shown for every party as a bar plot and the average as a vertical dashed line. A: Survey carried out in January 2016 (just after the elections of December 2015). B: Survey carried out in May 2019 (just after the elections of April 2019).
	}
	\label{fig:party_loc_survey}
\end{figure}

The Spanish Center of Sociological Research \cite{cis_data} (Centro de Investigaciones Sociol\'ogicas - CIS) carries out periodic surveys in which they ask a representative sample of the Spanish population to locate the main political parties in the left/right ideological scale. For this work we have used the studies 01-2016 3124 (question 19), carried out in January 2016 and 05-2019 3247 (question 25), carried out in May 2019. In Figure \ref{fig:party_loc_survey} we show the distribution of the answers. The data is provided already binned by CIS in the following intervals: Left (1-2), (3,4), (5,6), (7,8), (9-10) Right. 




\clearpage

\clearpage
\printbibliography